\documentclass[12pt,nofootinbib,preprint,superscriptaddress]{revtex4}
\pdfoutput=1

\usepackage{amsmath}
\usepackage{amssymb}
\usepackage{anysize}
\usepackage{bold-extra}
\usepackage{xcolor}
\usepackage{enumerate}
\usepackage{fancyhdr}
\usepackage{graphicx}
\usepackage[linktocpage,colorlinks,urlcolor=blue]{hyperref}
\usepackage{multirow}
\usepackage[final]{pdfpages}
\usepackage{rotating}
\usepackage{textcomp}
\usepackage{url}
\usepackage{verbatim}
\usepackage{wrapfig}
\usepackage{slashed}
\usepackage{subfigure}

\usepackage{amsfonts}
\usepackage{comment}
\usepackage{epigraph}
\usepackage[utf8]{inputenc}
\usepackage{slashed}
\usepackage{bbm}
\usepackage{cancel}
\usepackage{epsf, array, color}
\newcommand{\La}{\mathcal{L}}

\begin{document}

\title{\boldmath NLO Effects in EFT Fits to $W^+W^-$ Production at the LHC }

\author{Julien Baglio}
\email{julien.baglio@uni-tuebingen.de}
\affiliation{Institute for Theoretical Physics, University of
  T\"ubingen, Auf der Morgenstelle 14, 72076 T\"ubingen, Germany}

\author{Sally Dawson}
\email{dawson@bnl.gov}
\affiliation{Department of Physics, Brookhaven National Laboratory, Upton, N.Y., 11973~ U.S.A.}

\author{Ian M. Lewis}
\email{ian.lewis@ku.edu}
\affiliation{Department of Physics and Astronomy, University of Kansas, Lawrence, Kansas, 66045~ U.S.A.}

\begin{abstract}
We study the impact of anomalous gauge boson and fermion couplings on
the production of $W^+W^-$ pairs at potential future LHC upgrades and
estimate the sensitivity at $\sqrt{S}=14$~TeV with $3~ab^{-1}$ and
$\sqrt{S}=27$~TeV with $15~ab^{-1}$.  A general technique for
including NLO QCD effects in effective field theory (EFT)  fits to
kinematic distributions is presented, and numerical results are  given
for $\sqrt{S}=13$~TeV $W^+W^-$  production. Our method allows  fits
to anomalous couplings at NLO accuracy in any EFT basis and has been
implemented in a publicly available version of the {\tt POWHEG-BOX}.
Analytic expressions for the $K$-factors relevant for $13$~TeV total
cross sections are given for the HISZ and Warsaw EFT bases and
differential $K$-factors can be obtained using the supplemental
material. Our study demonstrates the necessity of including anomalous
$Z$- fermion couplings in the extraction of limits on anomalous
3-gauge-boson couplings.

\end{abstract}

\maketitle

\section{Introduction}

With the discovery of the Higgs boson, and subsequent measurements of
its properties at the LHC, the general features of the Standard Model
(SM) electroweak theory have been confirmed
experimentally~\cite{Dawson:2018dcd}. Measurements of the Higgs
couplings and production rates agree with SM predictions at the
$10-20\%$
level~\cite{Khachatryan:2016vau,Butter:2016cvz,Aaboud:2017vzb,Aaboud:2018xdt,Sirunyan:2018koj}
and there is no indication of the existence of any new particle at the
TeV scale. Going forward, the task is to make comparisons between
theory and data at the few percent level. This requires not only
high-luminosity LHC running, but also improved theoretical
calculations. Furthermore, Higgs physics cannot be studied in
isolation, but must be examined in the context of the entire set of SM
interactions.

$W^+W^-$ pair production is an example of a process
whose properties are highly restricted by LEP
measurements~\cite{Schael:2013ita}, yet still provides relevant
information about the Higgs sector from LHC data. The  production of
$W^+W^-$  pairs  provides a sensitive test of the electroweak gauge
structure, since in the SM there are delicate cancellations between
contributions from $s-$channel $\gamma$ and $Z$ exchange and $t-$
channel fermion exchange that maintain perturbative unitarity.
Deviations from the  form of the  $\gamma W^+W^-$  and  $Z W^+W^-$
vertices predicted by the SM spoil the cancellations  between
contributions that enforce unitarity. These deviations have been
studied for decades~\cite{Duncan:1985vj,Hagiwara:1986vm}, while the
importance of non-SM fermion-$Z$ interactions in  ensuring unitarity
in $W^+W^-$ pair production has only recently been
realized~\cite{Zhang:2016zsp,Baglio:2017bfe,Alves:2018nof,Grojean:2018dqj}.

Beyond the Standard Model (BSM)  physics effects in $W^+W^-$
production can be studied using effective Lagrangian techniques where
the new physics is parameterized as an operator expansion in inverse
powers of a high scale, $\Lambda$,
\begin{equation}
L_{\rm SMEFT}=
 L_{\rm SM}+\sum_{i,n}{C_i^{(n)}\over\Lambda^{n-4}}O_i^{(n)}+\ldots\,
\label{eq:smeft}
\end{equation}
where $O_i^{(n)}$ has mass dimension-$n$ and $L_{\rm SM}$ contains
the complete SM Lagrangian.  The subscript SMEFT denotes the SM Effective Field Theory where the
Higgs is assumed to be part of an $SU(2)$ doublet.
This approach further assumes that there are no new light degrees of
freedom. Neglecting flavor, there are 59 possible operators at
dimension-6~\cite{Buchmuller:1985jz,Grzadkowski:2010es}, but only a
small subset of these  contribute to $W^+W^-$ production. 
At high energies, the longitudinally polarized contributions to
$W^+W^-$ grow with energy faster than the SM contributions in the
presence of BSM physics. This implies that the LHC can put strong
constraints on the coefficients of the dimension-$6$ SMEFT operators.

We consistently work with a dimension-$6$ Lagrangian.  At
dimension-$8$, there are many other possible operators, not only
modifying the triple-gauge boson interactions but also new $4-$point
$ggW^+W^-$ interactions~\cite{Bellm:2016cks}. These operators
contribute at tree level at order ${\cal O}(1/\Lambda^4)$ in the EFT
and must be included in a study of the dimension-$8$
Lagrangian. However, the study of dimension-$8$ operators is beyond
the scope of this work, so that we do not consider operators modifying
the partonic cross section $gg\to W^+W^-$.

The effects of new physics contributions to  $W^+W^-$ gauge boson pair
production can be expected to be of the same order of magnitude as QCD
corrections, and so these contributions must be included when
extracting limits on new physics. QCD effects in the effective field
theory can also change the dependence of the experimental kinematic
distributions on the coefficients of Eq.~(\ref{eq:smeft}). The SM QCD
corrections to $W^+W^-$ pair production are known up to
NNLO~\cite{Gehrmann:2014fva,Grazzini:2016ctr}, including the effects
of a jet veto~\cite{Dawson:2016ysj,Hamilton:2016bfu}, and the
electroweak corrections exist at NLO~\cite{Bierweiler:2013dja,Baglio:2013toa,Biedermann:2016guo}.
The SM and dimension-6 $gg$ initial state contributions are formally NNLO and are not
included, although at $14$~TeV they increase the cross section by
roughly $10\%$ (see e.g. refs.~\cite{Campbell:2011bn,Chiesa:2018lcs}).
 We perform an analysis including QCD
 corrections~\cite{Dixon:1998py,Dixon:1999di} at NLO in the SMEFT,
 along with modifications of both the 3-gauge-boson and fermion
 couplings, extending our previous study~\cite{Baglio:2017bfe} by
 including the leptonic decays of the $W$'s. The effects of anomalous
 3-gauge boson couplings exist in the {\tt POWHEG-BOX}
 framework~\cite{Melia:2011tj,Nason:2013ydw}, and we add the
 additional contributions from anomalous fermion couplings. This
 public tool can be found  at  \url{http://powhegbox.mib.infn.it} and
 can be used to perform fits to anomalous couplings including NLO QCD
 and showering effects.

In Section~\ref{sec:basics}, we review the basics of the effective
field theory framework for $W$ pair production and discuss the effects
of  NLO QCD in the SMEFT.
The determination of NLO QCD effects in a theory with anomalous
couplings has typically been done on a case by case basis, or
alternatively by allowing one SMEFT coupling at a time to vary. In
Section~\ref{sec:prim}, we present a general method for deriving NLO
expressions for the total cross section and for distributions in an
SMEFT in terms of  a fixed number of sub-amplitudes, which we term
"primitive cross sections".  These results can be used to obtain
either total or differential $K$-factors in any SMEFT basis.   
In Section~\ref{sec:fp}, we compare projections for the measurements
of anomalous 3-gauge-boson couplings at the high-luminosity LHC with
those of a future $27$~TeV collider and demonstrate the critical
importance of including anomalous fermion couplings in the fits
 (see
also ref.~\cite{Biekotter:2018jzu} for SMEFT projections in a global
fit at a 27 TeV hadron collider).
We provide numerical results in Section~\ref{sec:13tev} for
$K$-factors for the leading lepton $p_T$ ($p_T^{\ell,lead}$) and
$m_{ll}$ distributions for  $W^+W^-\rightarrow e^\pm
\mu^{\mp}\nu{\overline{\nu}}$ at $13$~TeV as an illustration of our
technique, along with analytic results for the total cross section, as
functions of arbitrary EFT coefficients.

\section{Basics}
\label{sec:basics}
\subsection{Effective Gauge and Fermion Interactions}
Assuming CP conservation, the most general Lorentz invariant $3-$gauge
boson couplings are~\cite{Gaemers:1978hg,Hagiwara:1986vm},
\begin{eqnarray}
 \La_{V}=
-ig_{WWV}\biggl[g_1^V\left(W^+_{\mu\nu}W^{-\mu}V^\nu-W_{\mu\nu}^-W^{+\mu}V^\nu\right)+\kappa^VW^+_\mu
            W^-_\nu V^{\mu\nu}+\frac{\lambda^V}{M^2_W}W^+_{\rho\mu}{W^{-\mu}}_\nu V^{\nu\rho}\biggr],
\label{eq:lagdef}
\end{eqnarray}  
where $V=\gamma, Z$, $g_{WW\gamma}=e$ and $g_{WWZ}=g \cos\theta_W$,
 with $\theta_W^{}$ being the weak mixing angle, ($s_W^{} \equiv \sin\theta_W^{}$, $c_W^{} \equiv
\cos\theta_W^{}$). The fields in Eq.(\ref{eq:lagdef}) are the
canonically normalized mass eigenstate fields. 
We define $g_1^V = 1+\delta g_1^V$, $\kappa_{}^V=
1+\delta\kappa_{}^V$ and in the SM  $\delta g_1^V = \delta\kappa_{}^V
= \lambda_{}^V = 0$.  Gauge invariance requires
$\delta g_1^\gamma = 0$.

The effective couplings of quarks to gauge fields can be written as\footnote{We assume SM gauge couplings to leptons, since these couplings
are highly restricted by LEP data.}, (assuming
no new tensor structures),
\begin{eqnarray}
  \La&\equiv &g_ZZ_\mu\biggl[g_L^{Zq}+\delta g_{L}^{Zq}\biggr]
  {\overline q}_L\gamma_\mu q_L\
 +g_ZZ_\mu\biggl[g_R^{Zq}+\delta g_{R}^{Zq}\biggr]
  {\overline q}_R\gamma_\mu q_R\nonumber \\
  &&+{g\over \sqrt{2}}\biggl\{W_\mu\biggl[(1+\delta g_{L}^W){\overline u}_L\gamma_\mu d_L
  +\delta g_R^W
  {\overline u}_R\gamma_\mu d_R\biggr] +h.c.\biggr\}\, .
  \label{eq:dgdef}
  \end{eqnarray}
 $g_Z=e/(c_W^{}s_W^{})= g/c_W$, $Q_q$ is the electric
 charge of the quarks, and $q$ denotes up-type or down-type quarks. We assume the anomalous fermion
 couplings, $\delta g_{L,R}^{Zq}$,  along with the anomalous $W$ -fermion couplings are flavor independent.   We also neglect
 CKM mixing.
 The SM quark couplings are:
\begin{eqnarray}
g_R^{Zq}&=&-s_W^2 Q_q\quad{\rm and}\quad g_L^{Zq}=T_3^q -s_W^2 Q_q,
\end{eqnarray}
with $T_3^q=\pm \displaystyle \frac{1}{2}$.   $SU(2)$ invariance implies,
\begin{eqnarray}
\delta g_L^W&=&\delta g_L^{Zu}-\delta g_L^{Zd},
\nonumber \\
\delta g_1^Z&=& \delta \kappa_{}^Z+{s_W^2\over c_W^2}\delta \kappa_{}^\gamma,
\nonumber \\
\lambda_{}^\gamma &=& \lambda_{}^Z\, .
\label{eq:su2rel}
\end{eqnarray}
This framework leads
to $7$ unknown parameters: $\delta g_1^Z,~ \delta \kappa_Z,~\lambda_Z, ~\delta g_L^{Zu},~\delta g_L^{Zd},~\delta g_R^{Zu}$
and $~\delta g_R^{Zd}$\footnote{We neglect possible anomalous right-handed $W$-quark couplings, since they
are suppressed by small Yukawa couplings in an MFV framework.}.

At high energy scales the dominant contributions to $W^+W^-$ come from longitudinally polarized $W's$.
Keeping only the terms linear in the anomalous couplings, the amplitudes ${{\cal
    A}}_{rr'\lambda\lambda'}$ for ${\overline {q}}_r
q_{r^\prime}\rightarrow W^+_\lambda W^{-}_{\lambda '}$, where
$r,r',\lambda,\lambda'$ label the respective particle helicities,
have the high energy
limits~\cite{Gaemers:1978hg,Hagiwara:1986vm,Azatov:2016sqh,Falkowski:2016cxu,Baglio:2017bfe,Alves:2018nof,Grojean:2018dqj},
\begin{eqnarray}
  {{\cal A}}_{+-00}&\rightarrow&
                        {g^2 s\over 2 M_W^2}\sin\theta\biggl\{
                        \delta\kappa_{}^Z\biggl( s_W^2 Q_q^{} -
                        T_3^{q}\biggr) - s_W^2 Q_q^{}
                        \delta\kappa_{}^\gamma - \delta g_{L}^{Zq} + 2
                        T_3^{q} \delta g_L^W\biggr\},\nonumber\\
{{\cal A}}_{-+00}&\rightarrow &
                       {g^2 s\over 2 M_W^2}\sin\theta\biggl\{
                       s_W^2 Q_q\biggl(\delta\kappa_{}^\gamma-\delta
                       \kappa_{}^Z\biggr) +\delta g_{R}^{Zq} \biggr\}\, ,
 \label{eq:longlims}                    
\end{eqnarray}
where  $\sqrt{s}$ is the partonic sub-energy.
From Eq.(\ref{eq:longlims}), it is clear that the longitudinal polarizations do not depend on the full range of $7$ anomalous
couplings, but on 4 linear combinations when $u$ and $d$ contributions are included.  Note that the dependence on $\lambda_Z$ is subleading in $s$.
The transverse polarizations have a weaker dependence  on the energy scale and different dependences on the
anomalous couplings.   

The Lagrangians of Eqs.(\ref{eq:lagdef}) and (\ref{eq:dgdef}) can be
mapped onto the effective Lagrangian of Eq.(\ref{eq:smeft}). For future convenience, we consider the mapping
to the Warsaw basis~\cite{Grzadkowski:2010es,Dedes:2017zog} and the HISZ basis~\cite{Hagiwara:1986vm}.  
In the Warsaw basis~\cite{Grzadkowski:2010es} , the dimension-$6$ operators
relevant for our analysis are,
\begin{eqnarray}
L_{WARSAW}&=&
{C_{3W}\over\Lambda^2} \epsilon^{abc} W_\mu^{a\nu}W_\nu^{b\rho}W_\rho^{c\mu}
+{C_{HD}\over\Lambda^2}\mid \Phi^\dagger (D_\mu \Phi)\mid^2
+{C_{HWB}\over\Lambda^2} \Phi^\dagger\sigma^a\Phi W^a_{\mu\nu}B^{\mu\nu}\nonumber \\
&&+{C_{Hf}^{(3)}\over\Lambda^2}i\biggl(\Phi ^\dagger \overleftrightarrow {D}_\mu^a \Phi \biggr) {\overline f}_L\gamma^\mu \sigma^a f_L
+{C_{Hf}^{(1)}\over\Lambda^2}i\biggl(\Phi ^\dagger  \overleftrightarrow {D}_\mu \Phi \biggr){\overline f}_L \gamma^\mu  f_L
\nonumber \\ &&
+{C_{Hf}\over\Lambda^2}i\biggl(\Phi ^\dagger  \overleftrightarrow{D}_\mu \Phi \biggr){\overline q}_R\gamma^\mu  q_R
+{C_{Hud}\over\Lambda^2}i\biggl(\widetilde{\Phi}^\dagger D_\mu \Phi\biggr) \overline{u}_R\gamma^\mu d_R
\nonumber \\ && 
+{C_{ll}\over\Lambda^2}({\overline l}_L\gamma^\mu l_L)({\overline l}_L\gamma_\mu l_L)\, ,
\label{eq:ops}
\end{eqnarray}
where $f$ can be either a quark or a lepton,  $D_\mu \Phi=(\partial_\mu -i\,\frac{g}{2}\sigma^a
W^a_\mu-i\frac{g'}{2}B_\mu)\Phi$, $W^a_{\mu\nu}=\partial_\mu W^a_\nu
-\partial_\nu W^a_\mu+g\varepsilon^{abc}W^b_\mu W^c_\nu$,
$\Phi^\dagger \overleftrightarrow{D}_\mu \Phi=\Phi^\dagger D_\mu
\Phi-(D_\mu \Phi^\dagger)\Phi$, and $\Phi^\dagger
\overleftrightarrow{D}^a_\mu \Phi=\Phi^\dagger D_\mu
\sigma^a\Phi-(D_\mu \Phi^\dagger)\sigma^a\Phi$. $\Phi$ stands for the
Higgs doublet field with a vacuum expectation value $\langle\Phi\rangle =
(0,v/\sqrt{2})^{\rm T}$.

In the HISZ basis, the fermion couplings are unchanged, while the 3-gauge-boson couplings are,
\begin{eqnarray}
L_{HISZ}&=&{f_W\over\Lambda^2}(D_\mu\Phi)^\dagger{\hat{W}}^{\mu\nu}D_\nu \Phi
+
{f_B\over\Lambda^2} (D_\mu\Phi)^\dagger{\hat{B}}^{\mu\nu}D_\nu \Phi
+
{f_{WWW}\over\Lambda^2}Tr\biggl({\hat {W}}_{\mu\nu}{\hat{W}}^{\nu\rho}{\hat{W}}_\rho^\mu
\biggr)\, ,
\label{eq:hiszbas}
\end{eqnarray}
where ${\hat{W}}^{\mu\nu}=i{g\over 2}\sigma^aW^{a,\mu\nu}$ and  
${\hat{B}}^{\mu\nu}=i{g^\prime\over 2}B^{\mu\nu}$.
Expressions for the anomalous 3-gauge-boson couplings  are given in Table \ref{tab:rgb}  and for the anomalous fermion
couplings in Table ~\ref{tab:ferm}.\footnote{We neglect dipole operators since
they do not interfere with the SM contributions.} 

\begin{table}
\centering
\begin{tabular}{|c||c|c|}
\hline
& Warsaw  Basis 
& HISZ\\
\hline\hline
$\delta g_1^Z$ & $\frac{v^2}{\Lambda^2}\frac{1}{c_W^2-s_W^2}\left(\frac{s_W^{}}{c_W^{}}C_{HWB}^{} + \frac14 C_{HD}^{} +\delta v\right)$&
${M_Z^2\over 2\Lambda^2}f_W$\\
\hline
$\delta \kappa_{}^Z$ &$\frac{v^2}{\Lambda^2}\frac{1}{c_W^2-s_W^2}\left(2 s_W c_W C_{HWB}^{} + \frac14 C_{HD}^{} +\delta v\right)$ & 
${M_Z^2\over 2\Lambda^2}(c_W^2f_W-s_W^2f_B)$\\
\hline
$\delta \kappa_{}^\gamma$ & $-\frac{v^2}{\Lambda^2}\frac{c_W^{}}{s_W^{}}C_{HWB}^{}$ &
${M_W^2\over 2\Lambda^2}(f_W+f_B)$ \\
\hline
$\lambda_{}^\gamma$ &$\frac{v}{\Lambda^2} 3 M_W^{} C_{3W}$ & 
${3g^2M_W^2\over 4\Lambda^2}f_{WWW}$\\
\hline
$\lambda_{}^Z$ & $ \frac{v}{\Lambda^2} 3 M_W^{} C_{3W}$& 
${3g^2 M_W^2\over 4\Lambda^2}f_{WWW}$ \\
\hline
\hline
\end{tabular}
\caption{Anomalous 3-gauge-boson couplings in the
  Warsaw~\cite{Grzadkowski:2010es} and
  HISZ~\cite{Hagiwara:1986vm}.~$\delta v$ is given in Table
  \ref{tab:ferm}.}
\label{tab:rgb}
\end{table}

\begin{table}
\centering
\begin{tabular}{|c||c|}
\hline
& Warsaw  Basis \\
\hline\hline
$\delta g_L^{Zu}$&$ -\frac{v^2}{2\Lambda^2}\left(C_{Hq}^{(1)}-C_{Hq}^{(3)}\right) + \frac12 \delta g_Z + \frac23\left(\delta s_W^2 - s_W^2 \delta g_Z^{}\right)$ \\
\hline
$\delta g_L^{Zd}$& $-\frac{v^2}{2\Lambda^2}\left(C_{Hq}^{(1)}+C_{Hq}^{(3)}\right) - \frac12 \delta g_Z - \frac13\left(\delta s_W^2 - s_W^2 \delta g_Z^{}\right)$\\
\hline
$\delta g_R^{Zu}$&$-\frac{v^2}{2\Lambda^2} C_{Hu} + \frac23\left(\delta s_W^2 - s_W^2\delta g_Z^{}\right)$ \\
\hline
$\delta g_R^{Zd}$& $-\frac{v^2}{2\Lambda^2} C_{Hd} - \frac13\left(\delta s_W^2 - s_W^2\delta g_Z^{}\right)$\\
\hline
$\delta g_L^W$&  $ \frac{v^2}{\Lambda^2}C_{Hq}^{(3)} + c_W^2\delta g_Z^{} + \delta s_W^2$\\
\hline
$\delta g_Z$ & $-\frac{v^2}{\Lambda^2}\left(\delta v +\frac14 C_{HD}^{}\right)$ \\ 
\hline
$\delta v$  &$C_{Hl}^{(3)} - \frac12 C_{ll}^{}$ \\
\hline
$\delta s_W^2$  & $ -\frac{v^2}{\Lambda^2} \frac{s_W^{} c_W^{}}{c_W^2-s_W^2}\left[2 s_W^{} c_W^{}\left(\delta v + \frac14 C_{HD}^{}\right) + C_{HWB}^{}\right]$ \\
\hline
\hline
\end{tabular}
\caption{Anomalous fermion couplings in the Warsaw~\cite{Grzadkowski:2010es} basis.}
\label{tab:ferm}
\end{table}

\subsection{Primitive Cross Sections}
\label{sec:prim}

We want to compute differential and total cross sections for a hadronic scattering process at NLO QCD for arbitrary
anomalous couplings.  Since these calculations can be numerically intensive, it is desirable not to have to repeat the
calculation over and over again for different values of the anomalous couplings.  Here we discuss a technique for generating
results in terms of a set of primitive cross sections which need to be calculated only once for a given process and set of cuts. 
The primitive cross sections  can be reweighted to 
 allow for rapid scans  over the anomalous couplings at NLO order.  

Consider an arbitrary  differential cross section $d\sigma^n({\vec{C}})$ that
is calculated to ${\cal O}({ \Lambda^{-2n}})$.  It depends  on $m$ EFT coefficients $\vec{C}=(C_1,C_2,\ldots,C_m)$ and
the relevant momenta, ${\vec{p}}$.  We assume $C_i\sim {\cal{O}}({\Lambda^{-2}})$. It is important to note that in general $d\sigma^4$ is the cross section to order $\mathcal{O}({\Lambda^{-4}})$, but is not the amplitude-squared.   When both $Z-$ fermion and three gauge boson (3GB) couplings are non-zero,
the amplitude-squared contains terms up to ${\cal{O}}({\Lambda^{-8}})$.

Calculating the cross section to 
${\cal O}({\Lambda^{-2}})$, 
\begin{equation}
d\sigma^1({\vec{C}})\equiv d\sigma_{SM}(1-\Sigma_{i=1}^{m} C_i )+\Sigma_{i=1}^m C_i\, d\sigma(1; {\vec{R}}_i)\, ,
\label{eq:master2}
\end{equation}
where  $\vec{R}_i$ are $m$- dimensional vectors with 
$\vec{R}_1=(1,0,0.....0)$, $\vec{R}_2=(0,1,0....)$, $\vec{R}_m=(0,0....1)$, etc.  
The primitive cross section $d\sigma(n;\vec{R}_i)$ is the cross section obtained to arbitrary order
${\cal{O}}({\Lambda^{-2n}})$ when $C_i=1$ and all other $C_j=0$, $j\ne i$:
\begin{eqnarray}
d\sigma(n;\vec{R}_i)\equiv d\sigma^n(\vec{C}=\vec{R}_i).\label{eq:prim1}
\end{eqnarray}
The SM cross section with $\vec{C}=0$ is $d\sigma_{SM}$.  
For the process $pp\rightarrow
W^+W^-$ under consideration here, there are 7 primitive cross sections, $d\sigma(1;\vec{R}_i)$, when considering both anomalous 3GB couplings and  
anomalous $Z-$fermion 
couplings, while there are only 3 when  the fermion couplings take their SM values. Eq.(\ref{eq:master2}) holds bin by
bin for differential rates and also for the total rate. 

The procedure becomes significantly more laborious if the cross section is computed to ${\cal{O}}({\Lambda^{-4}})$.
We  define an $m$-dimensional  vectors ${\vec M}_{ij}$, with
$\vec{M}_{12}=(1,1,0,...0)$, $\vec{M}_{13}=(1,0,1....0)$, $\vec{M}_{23}=(0,1,1,....)$, etc.
To ${\cal O}({\Lambda^{-4}})$, the cross section decomposes into the primitive cross sections as, 
\begin{eqnarray}
d\sigma^2({\vec{C}})&=&d\sigma_{SM}(1-\Sigma_{i=1}^{m} C_i )+\Sigma_i C_i\, d\sigma(1; {\vec{R}}_i)
+\Sigma_{i=1}^m C_i^2\,\biggl(d\sigma(2;{\vec{R}}_i)-d\sigma(1;{\vec{R}}_i)\biggr)
\nonumber \\
&&+\Sigma_{i>j=1}^m C_iC_j\biggl[ d\sigma(2;\vec{M}_{ij})-d\sigma(2;{\vec{R}}_i)-d\sigma(2;{\vec{R}}_j)+d\sigma_{SM}\biggr]\, .
\label{eq:fits}
\end{eqnarray}
  Evaluating the cross section with $2$ non-zero $C_i$ coefficients, $C_i=1$, $C_j=1$, and $C_{k\ell}=0$ if $j\neq i$ or $k\neq \ell$, to arbitrary order $\mathcal{O}(\Lambda^{-2n})$ yields the
primitive cross sections 
\begin{eqnarray}
d\sigma(n;\vec{M}_{ij})\equiv d\sigma^n(\vec{C}=\vec{M}_{ij}).\label{eq:prim2}
\end{eqnarray}
We can calculate the primitive cross sections, $d\sigma(n;{\vec{R}_i})$ and $d\sigma(n;{\vec{M}_{ij}})$ once 
and then apply Eq.(\ref{eq:fits}) to get the  general result  for
arbitrary anomalous couplings. Eqs.(\ref{eq:master2}) and (\ref{eq:fits}) hold separately
for LO and NLO corrected rates.   For  $pp\rightarrow
W^+W^-$, there are 35 primitive cross sections at ${\cal{O}}({\Lambda^{-4}})$ with both anomalous 3GB couplings and  anomalous fermion 
couplings non-zero.

Suppose we calculate the primitive cross sections in terms of a set of  EFT parameters, $\vec{C}=(C_1,\ldots,C_m)$, 
but we want the results in terms
of a different EFT basis, $\vec{C}^\prime=(C_1^\prime,\ldots, C_m^\prime)$.   The procedure is straightforward. 
We begin by considering  $2$ anomalous couplings, $m=2$,  and the $\mathcal{O}(\Lambda^{-2})$ case.  The physical rate must be independent 
of the basis choice and 
using the  master formula of Eq.(\ref{eq:master2}) for the $2$ different bases,
\begin{eqnarray}
d\sigma^1 ({\vec{C}}) &=&d\sigma_{SM}(1-C_1-C_2)+C_1\,d\sigma(1;\vec{R}_1)+C_2\,d\sigma(1;\vec{R}_2)  \nonumber \\
d\sigma^1 ({\vec{C^\prime }})&=& d\sigma_{SM}(1-C_1^\prime-C_2^\prime)+C_1^\prime\,d\sigma^\prime (1;\vec{R}_1)+C_2^\prime\,d\sigma^\prime(1;\vec{R}_2).
\end{eqnarray}
The unprimed primitive cross sections are defined according to Eq.(\ref{eq:prim1}).   To order $\mathcal{O}(\Lambda^{-2n})$ the primed primitive cross sections $d\sigma^\prime(n,\vec{R}_i\prime)$ are defined with the vector $\vec{R}_i$ evaluated relative to the new basis $\vec{C}'$
\begin{eqnarray}
d\sigma^{\prime}(n;\vec{R}_i)\equiv d\sigma^n(\vec{C'}&=&\vec{R}_i).\label{eq:prim3}
\end{eqnarray}
We need the $d\sigma^\prime$'s in terms of the  already computed $d\sigma$'s so that we can avoid recalculating the cross sections.    
Including a complete basis of dimension-6 operators, the input parameters are related by a linear transformation,
\begin{equation}
\left( \begin{array}{c}
C_1\\
C_2\\
\end{array}\right)
=\alpha \left( \begin{array}{c}
C_1^\prime\\
C_2^\prime\\
\end{array}\right)\label{eq:2Dtrans}
\, ,
\end{equation}
where $\alpha$ is a $2\times2$ matrix.
The $d\sigma^\prime$ matrices are:
\begin{eqnarray}
d\sigma^\prime(1;\vec{R}_1)&=&d\sigma_{SM}(1-\alpha_{11}-\alpha_{21})+\alpha_{11}\,d\sigma(1;\vec{R}_1) +\alpha_{21}\,d\sigma(1;\vec{R}_2)\\
d\sigma^\prime(1;\vec{R}_2)&=&d\sigma_{SM}(1-\alpha_{12}-\alpha_{22})+\alpha_{12}\,d\sigma(1;\vec{R}_1) +\alpha_{22}\,d\sigma(2;\vec{R}_2)\, .
\label{eq:trans}
\end{eqnarray}

Now consider the general case of a change of EFT input basis.  Assume we have two minimum sets\footnote{All redundant
operators have been eliminated using the equations of motion.}
 of independent parameters $\vec{C}=(C_1,C_2,\cdots,C_m)$ and $\vec{C'}=(C'_1,C'_2,\cdots,C'_m)$ that can be related linearly,
\begin{eqnarray}
C_i=\sum_{j=1}^m \alpha_{ij}C_j^\prime,\quad C_i^\prime=\sum_{j=1}^m \alpha^{-1}_{ij}C_j,\label{eq:coups}\end{eqnarray}
where $\alpha^{-1}$ is the inverse matrix of $\alpha$ and $\alpha^{-1}_{ij}$ is its $\{i,j\}^{th}$ element. 

  Although the two parameter bases $C_i,C'_i$ must give identical results for physical quantities and, \footnote{We suppress possible momentum dependence.}
\begin{eqnarray}
\sigma^n\equiv\sigma^n(\vec{C})=\sigma^n(\vec{C'})\, ,
\end{eqnarray}
 the primitive cross sections are not the same since setting $C_i=1$ is not the same as  taking $C'_i=1$.  To $\mathcal{O}(\Lambda^{-2n})$, we define the primitive cross sections with two non-zero $C^\prime_j$ in the $\vec{C}'$ bases to be
\begin{eqnarray}
d\sigma^{\prime}(2;\vec{M}_{ij})\equiv d\sigma^n(\vec{C'}=\vec{M}_{ij}^\prime)\,\label{eq:prim4}
\end{eqnarray}
where $\vec{M}_{ij}$ evaluated relative to the $\vec{C}$ basis.  The primitive cross sections $d\sigma(n;\vec{R}_i)$, $d\sigma(n;\vec{M}_{ij})$, and $d\sigma^\prime(n;\vec{R}_j)$ are defined in Eqs.(\ref{eq:prim1}),~(\ref{eq:prim2}), and (\ref{eq:prim3}), respectively.  

The cross sections can now be expanded in terms of either set of parameters and primitive cross sections.  At $\mathcal{O}(\Lambda^{-2})$,  we have the cross sections\footnote{This applies for total cross sections, or bin by bin for differential cross sections.}
\begin{eqnarray}
d\sigma^1&=&d\sigma_{SM}+\sum_{i=1}^m C_i\left(d\sigma(1;\vec{R}_i)-d\sigma_{SM}\right)\nonumber\\
&=&d\sigma_{SM}+\sum_{i=1}^m C'_i\left(d\sigma'(1;\vec{R}_i)-d\sigma_{SM}\right)\label{eq:lam2},
\end{eqnarray}
where $\sigma_{SM}$ is the SM  cross section with $\vec{C}=\vec{C'}=\vec{0}$.  At order $\mathcal{O}(\Lambda^{-4})$,  we have
\begin{eqnarray}
d\sigma^2&=&d\sigma_{SM}+\sum_{i=1}^m C_i \left(d\sigma(1; \vec{R}_i)-d\sigma_{SM}\right)
+\sum_{i=1}^m C_i^2\left(d\sigma(2;\vec{R}_i)-d\sigma(1;\vec{R}_i)\right)\nonumber \\
&&+\Sigma_{i>j=1}^m C_iC_j\left( d\sigma(2;\vec{M}_{ij})-d\sigma(2;\vec{R}_i)-d\sigma(2;\vec{R}_j)+d\sigma_{SM}\right)\nonumber\\
&=&d\sigma_{SM}+\sum_{i=1}^m C'_i \left(d\sigma'(1; \vec{R}_i)-d\sigma_{SM}\right)
+\sum_{i=1}^m {C'}_i^2\left(d\sigma'(2;\vec{R}_i)-d\sigma'(1;\vec{R}_i)\right)\nonumber \\
&&+\Sigma_{i>j=1}^m C'_iC'_j\left( d\sigma'(2;\vec{M}_{ij})-d\sigma'(2;\vec{R}_i)-d\sigma'(2;\vec{R}_j)+d\sigma_{SM}\right)\, .
\label{eq:lam4}
\end{eqnarray}
Finally, we find the relationships between the primitive cross sections using Eqs.~(\ref{eq:coups},\ref{eq:lam2},\ref{eq:lam4}) that can be used to calculate cross sections in
terms of an arbitrary EFT basis:
\begin{eqnarray}
d\sigma^\prime(1;\vec{R}_i)&=&d\sigma_{SM}+\sum_{k=1}^m \alpha_{ki}\left(d\sigma(1;\vec{R}_k)-d\sigma_{SM}\right)\nonumber \\
d\sigma^\prime(2;\vec{R}_i)&=&d\sigma^\prime(1;\vec{R}_i)+\sum_{k=1}^m \alpha_{ki}^2\left(d\sigma(2;\vec{R}_k)-d\sigma(1;\vec{R}_k)\right)\nonumber \\
&+&\sum_{k>l=1}^m \alpha_{ki}\alpha_{li}\left(d\sigma(2;\vec{M}_{kl})-d\sigma(2;\vec{R}_k)-d\sigma(2;\vec{R}_l)+d\sigma_{SM}\right)\nonumber\\
d\sigma^\prime(2;\vec{M}_{ij})&=&d\sigma^\prime(1;\vec{R}_i)+d\sigma^\prime(1;\vec{R}_j)-d\sigma_{SM}\nonumber \\
&+&\sum_{k=1}^m\left(\alpha_{ki}+\alpha_{kj}\right)^2\left(d\sigma(2;\vec{R}_k)-d\sigma(1;\vec{R}_k)\right)\\
&+&\sum_{k>l=1}^m\left(\alpha_{ki}+\alpha_{kj}\right)
\left(\alpha_{li}+\alpha_{lj}\right)
( 
d\sigma(2;\vec{M}_{kl})
-d\sigma(2;\vec{R}_k)-d\sigma(2;\vec{R}_l)+d\sigma_{SM})\, ,\nonumber
\end{eqnarray}
or equivalently,
\begin{eqnarray}
d\sigma(1;\vec{R}_i)&=&d\sigma_{SM}+\sum_{k=1}^m \alpha^{-1}_{ki}\left(d\sigma^\prime(1;\vec{R}_k)-d\sigma_{SM}\right)\nonumber \\
d\sigma(2;\vec{R}_i)&=&d\sigma(1;\vec{R}_i)+\sum_{k=1}^m(\alpha_{ki}^{-1})^2\left(d\sigma^\prime(2;\vec{R}_k)-d\sigma^\prime(1;\vec{R}_k)\right)\nonumber \\
&+&\sum_{k>l=1}^m \alpha^{-1}_{ki}\alpha^{-1}_{li}\left(d\sigma^\prime(2;\vec{M}_{kl})-d\sigma^\prime(2;\vec{R}_k)-d\sigma^\prime(2;\vec{R}_l)+d\sigma_{SM}\right)\nonumber\\
d\sigma(2;\vec{M}_{ij})&=&d\sigma(1;\vec{R}_i)+d\sigma(1;\vec{R}_j)-d\sigma_{SM}
\nonumber \\
&+&\sum_{k=1}^m\left(\alpha^{-1}_{ki}+\alpha^{-1}_{kj}\right)^2\left(d\sigma^\prime(2;\vec{R}_k)-d\sigma^\prime(1;\vec{R}_k)\right)\nonumber\\
&+&\sum_{k>l=1}^m\left(\alpha^{-1}_{ki}+\alpha^{-1}_{kj}\right)\left(\alpha^{-1}_{li}+\alpha^{-1}_{lj}\right)\left( d\sigma^\prime(2;\vec{M}_{kl})-d\sigma^\prime(2;\vec{R}_k)-d\sigma^\prime(2;\vec{R}_l)+d\sigma_{SM}\right)\nonumber\, .
\end{eqnarray}
The above results  are found using
\begin{eqnarray}
C_iC_j=\sum_{k=1}^m \alpha_{ik}\alpha_{jk}{C'_k}^2+\sum_{k>l=1}^m\left(\alpha_{ik}\alpha_{jl}+\alpha_{il}\alpha_{jk}\right)C'_kC'_l,
\end{eqnarray}
which simplifies for $i=j$
\begin{eqnarray}
{C}_i^2=\sum_{k=1}^m \alpha^2_{ik}{C'}^2_k+2\sum_{k>l=1}^m \alpha_{ik}\alpha_{il}C'_kC'_l.
\end{eqnarray}
We illustrate the procedure and the utility of the results in this section by transforming from the Warsaw to HISZ basis in Sec. \ref{sec:13tev}.

\section{Future Projections}
\label{sec:fp}

In this section, we apply the results of the previous sections to
projecting allowed regions for anomalous 3-gauge-boson and $Z-$fermion
couplings at a high-luminosity LHC (HL-LHC) and a potential $27$~TeV
collider (HE-LHC) under various assumptions about the systematic
uncertainties We have extended the {\tt
  POWHEG-BOX-V2}~\cite{Nason:2004rx,Frixione:2007vw,Alioli:2010xd}
implementation of the NLO QCD corrected predictions for the process
$pp\to W^+W^-\to
4\ell$~\cite{Melia:2011tj,Nason:2013ydw}\footnote{Our
  extension is built upon an updated private version. We thank
  Giulia Zanderighi for it.}, that
contains only the 3-gauge-boson anomalous couplings, as originally
found in Refs.~\cite{Dixon:1998py,Dixon:1999di} and implemented also
in {\tt MCFM}~\cite{Campbell:1999ah}. In our implementation
in the {\tt POWHEG-BOX-V2}, we also include the anomalous Z-fermion
couplings, there is the option to choose the order of the
$\Lambda^{-2n}$ expansion and the results can be  generated using the
effective interactions of Eqs.(\ref{eq:lagdef}) and (\ref{eq:dgdef})
or the Warsaw basis coefficients of Eq.(\ref{eq:ops}). Our projections
assume the $\Lambda^{-4}$ expansion. Note that our extension works for
the case of different-flavor leptonic final states. In the case of
same-flavor charged leptons the contribution from $ZZ$ production as
well as the interference between $W^+W^-$ and $ZZ$ contributions
should be included.

We apply the basic cuts, 
\begin{eqnarray}
p_T^\ell>30~{\rm GeV},\,|\eta^\ell|<2.5,\,m_{\ell\ell}>10~{\rm
  GeV},\,\slashed{E}_T>20~{\rm GeV},
 \ell=e^\pm,\mu^\mp\, ,
 \label{eq:standcuts}
\end{eqnarray}
where $p_T^\ell$ is charged lepton transverse momentum,
  $\eta^\ell$ is charged lepton rapidity, $m_{\ell\ell}$ is the
  invariant mass of the two charged leptons, and $\slashed{E}_T$ is
  the missing energy of the event.  Since we do not include detector
  effects, in our case the missing transverse energy is the transverse
  energy of the two final state neutrinos.
We work at the parton level and veto jets  with
\begin{eqnarray}
p_T^{jet}>35~{\rm GeV},\,|\eta^{jet}|<4.5\, ,
\label{eq:jetcuts}
\end{eqnarray}
where $p_T^{jet}$ is the jet transverse momentum and $\eta^{jet}$ is
the jet rapidity.  We further assume a $50\%$ efficiency, in line with
the experimental results of Ref.~\cite{Aad:2016wpd}.  We use CT14QED
Inclusive PDFs~\cite{Schmidt:2015zda}, and take the
renormalization/factorization scales equal to $\displaystyle
{M_{WW}\over 2}$.

We assume a systematic uncertainty of $\delta_{syst}=16\%$ and find
the point where the systematic and statistical errors on the leading
lepton $p_T$ distribution are  roughly equal,
$\delta_{stat}\sim\delta_{sys}$. 
 At this point, the uncertainties are systematics dominated and increased statistics provide diminishing returns.  (In our figures, we investigate the impact of reducing the
systematic error to $\delta_{syst}=4\%$.)
 We define this point in terms of  the
integral over the leading lepton transverse momentum
$p_{T}^{\ell,lead}$ above a cut $p_{T,cut}$,
\begin{eqnarray}
\sigma(p_{T}^{\ell,lead} > p_{T,cut})=\int_{p_{T,cut}}
  \frac{d\sigma}{dp_{T}^{\ell,lead}}dp_{T}^{\ell,lead}\, .
\end{eqnarray}
The $p_T$ cut is defined to be the point where,
\begin{eqnarray}
&&\delta_{sys}=
   \delta_{stat}=\frac{1}{\sqrt{N}}=\frac{1}{\sqrt{L\sigma(p_{T}^{\ell,lead}>
   p_{T,cut})}}\nonumber \\
&\Rightarrow&\sigma(p_{T}^{\ell,lead}>p_{T,cut})=\frac{1}{L\delta^2_{sys}}.
\end{eqnarray}
$L$ is the integrated luminosity and $N$ is the number of events
passing the cuts. This corresponds to roughly $38$ events above the
cut. Using integrated luminosities of $3~fb^{-1}$ at $14$~TeV and
$15~fb^{-1}$ at $27$~TeV, we determine the $p_T$ cuts: 
\begin{eqnarray}
14~\text{TeV}:  
~ p_{T,cut}&=&750~GeV\nonumber \\
27~\text{TeV}:  
~p_{T,cut}&=&1350~GeV\, . 
\label{eq:ptcuts}
\end{eqnarray}
Retaining $\delta_{stat}=16\%$ with the corresponding $p_T$ cuts of
Eq.(\ref{eq:ptcuts}), we also consider the effect of reducing the
systematic error to $\delta_{sys}=4\%$.

\begin{figure}[h!]
  {\centering
    \includegraphics[width=0.45\textwidth,clip]{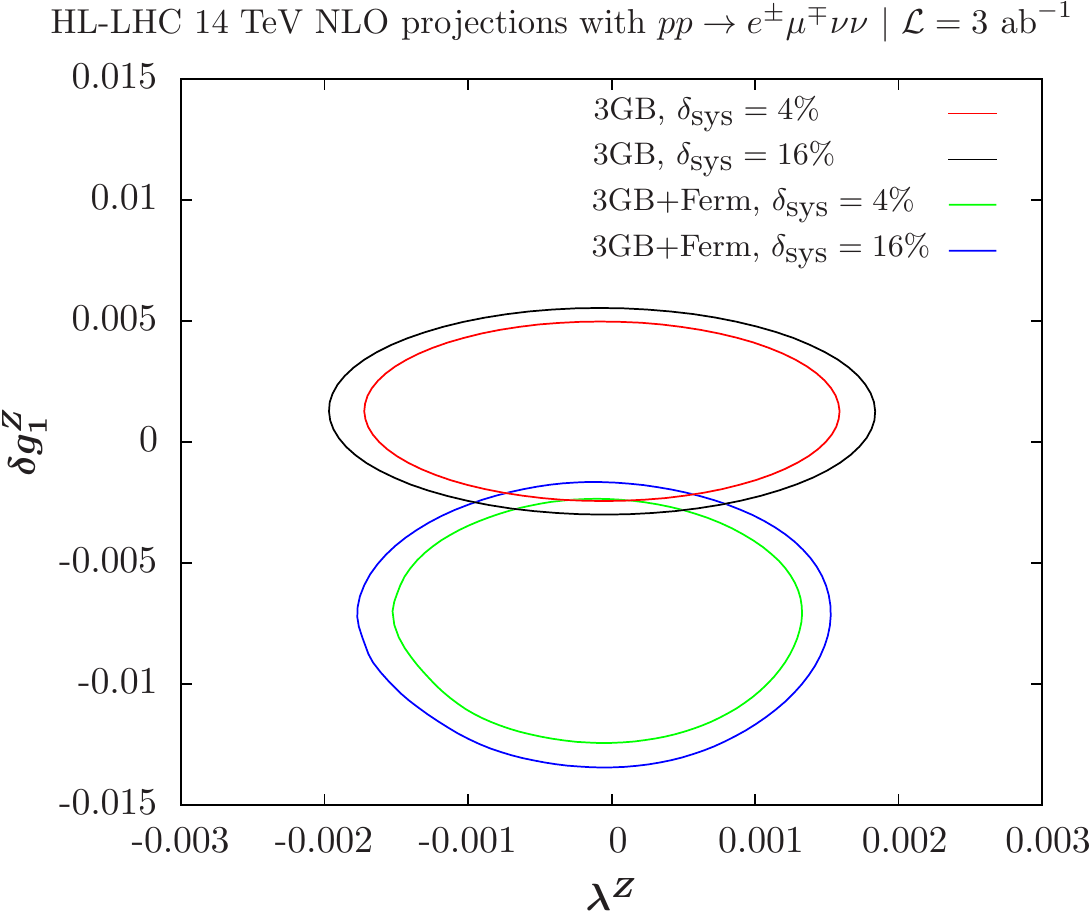}
    \includegraphics[width=0.45\textwidth,clip]{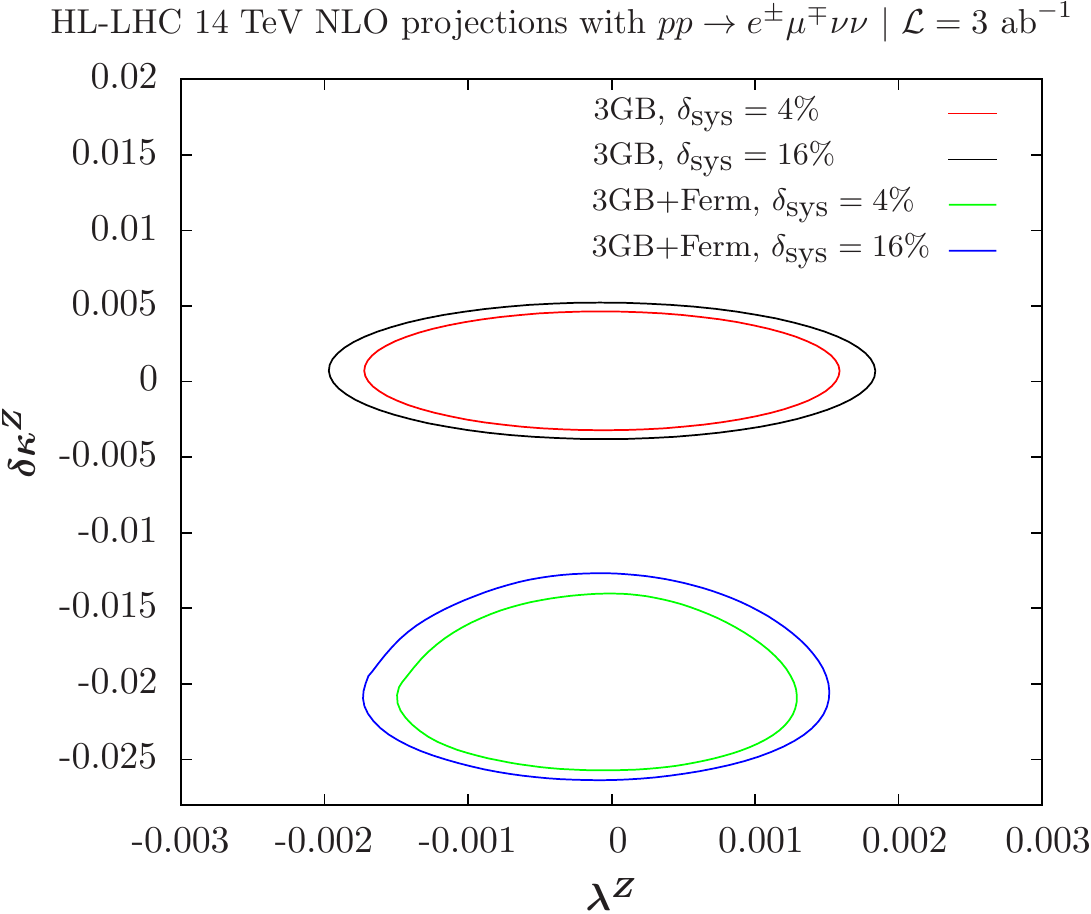}\bigskip\\
    \includegraphics[width=0.45\textwidth,clip]{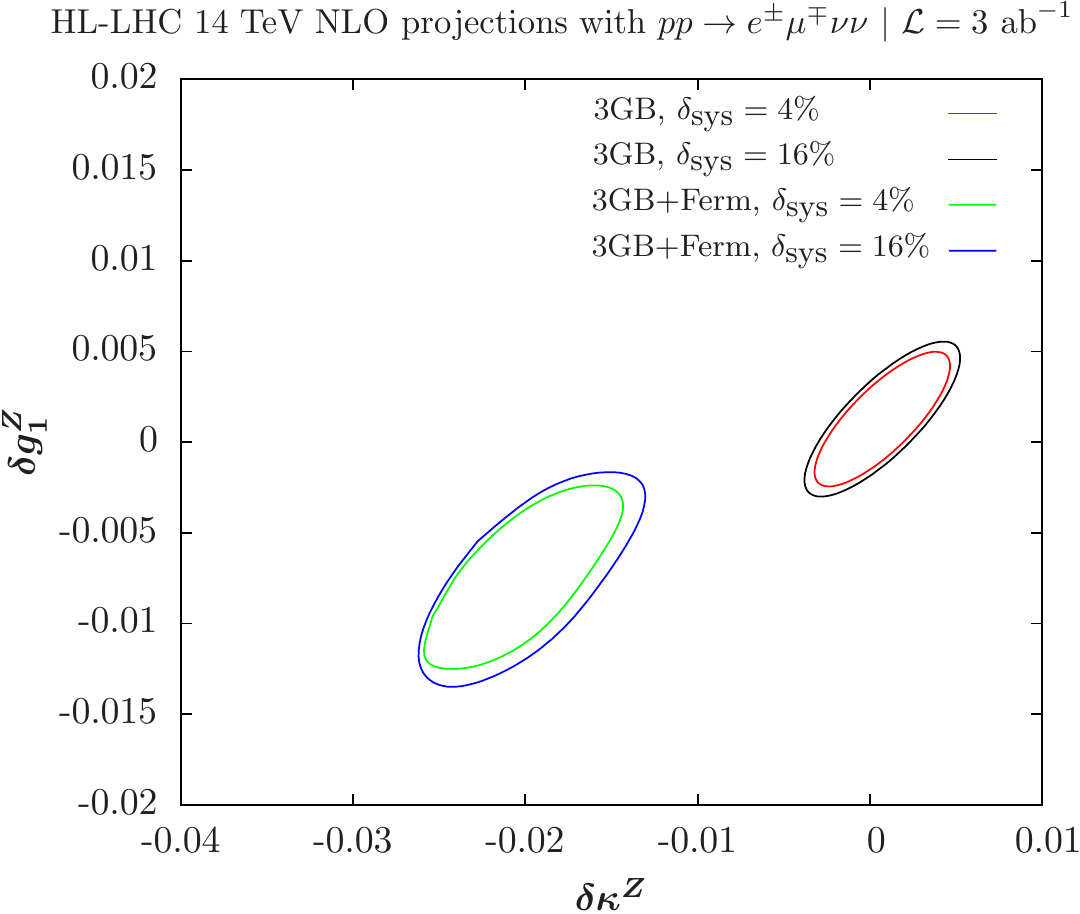}
  }
  \caption{Projections for the $14$~TeV HE-LHC with $3~ab^{-1}$. We
    take $p_{T,cut}=750$~GeV, corresponding to $\delta_{stat}=16\%$
    and consider $\delta_{sys}=16\%$ and $\delta_{sys}=4\%$. The
    curves labeled ``3GB'' have SM $Z-$fermion couplings (black and
    red), while the curves labeled ``3GB+Ferm'' allow the fermion
    couplings to vary within the $2\sigma$ region defined by
    Eq.(\ref{eq:leplim}) (blue and green).   The standard cuts given
    in Eqs.~(\ref{eq:standcuts}) and (\ref{eq:jetcuts}) are applied.}
  \label{fig:fit14}
\end{figure}

\begin{figure}[h!]
  {\centering
    \includegraphics[width=0.45\textwidth,clip]{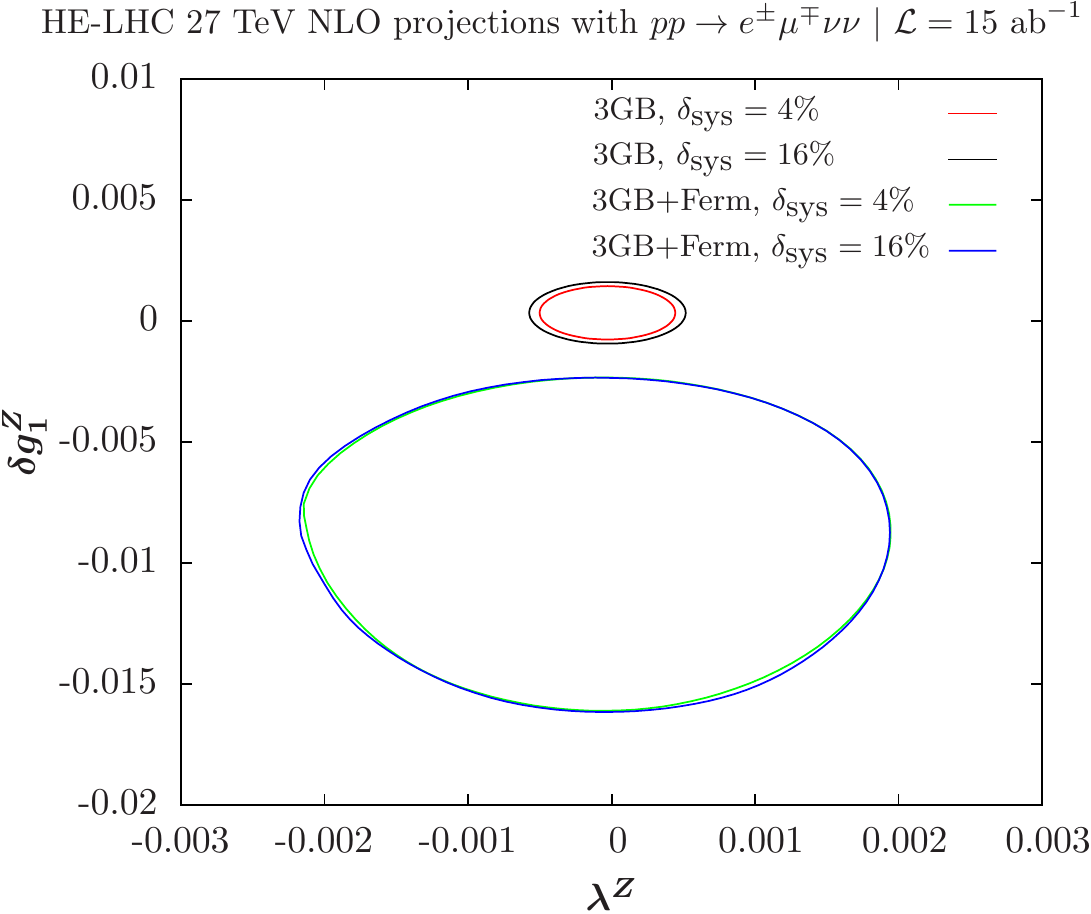}
    \includegraphics[width=0.45\textwidth,clip]{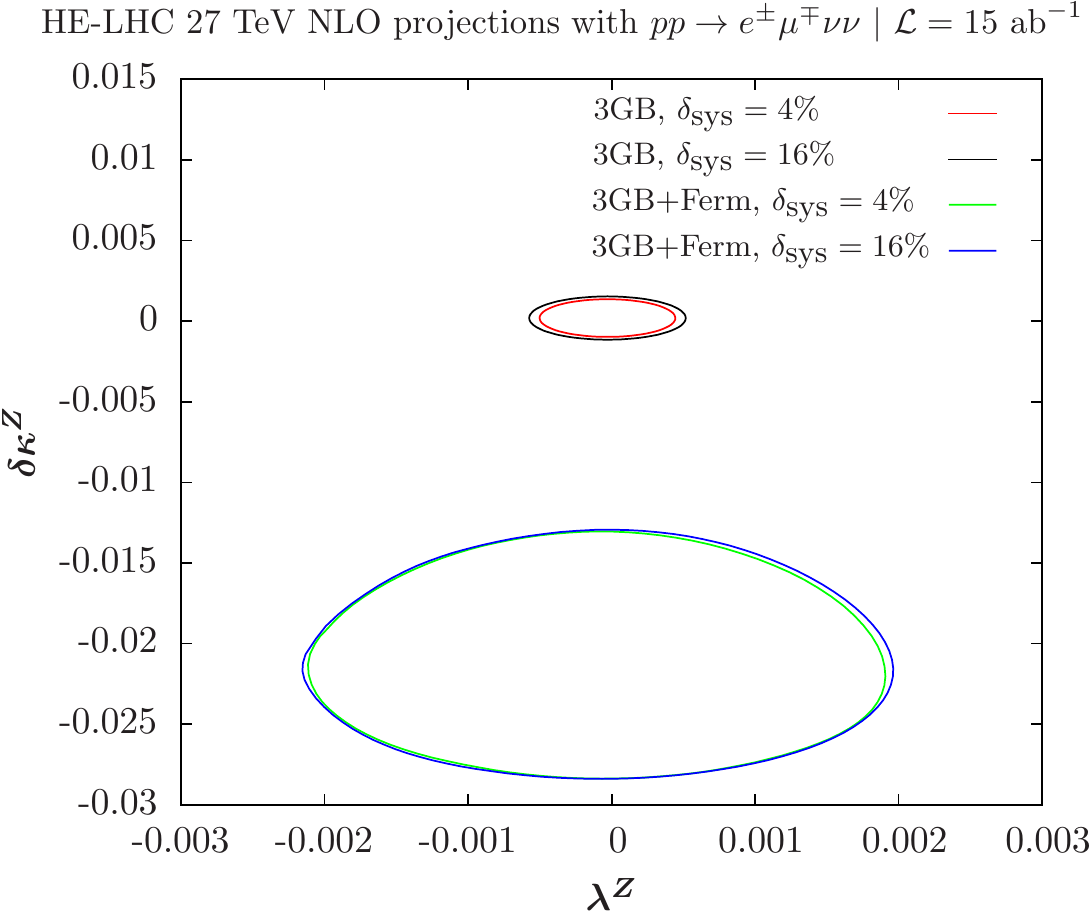}\bigskip\\
    \includegraphics[width=0.45\textwidth,clip]{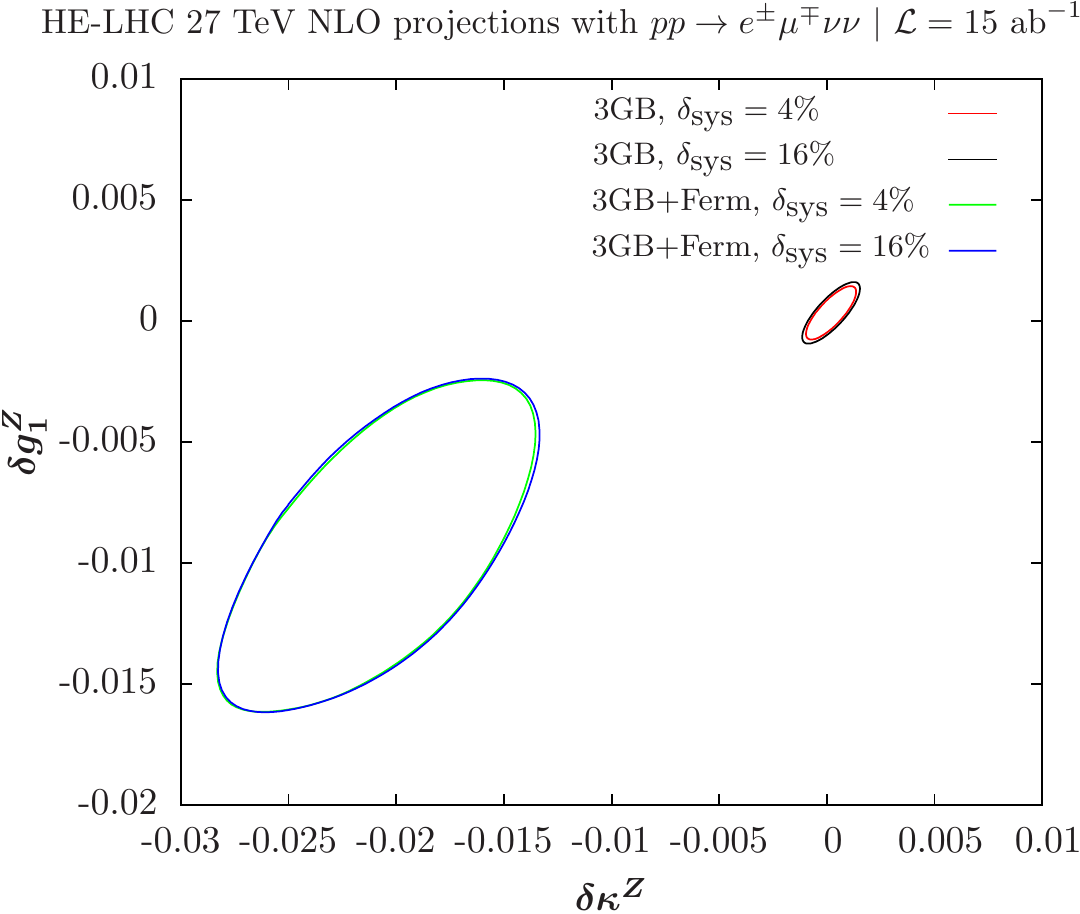}
  }
  \caption{Projections for the $27$~TeV HE-LHC with $15~ab^{-1}$.We
    take $p_{T,cut}=1350$~GeV, corresponding to $\delta_{stat}=16\%$
    and consider $\delta_{sys}=16\%$ and $\delta_{sys}=4\%$. The
    curves labeled ``3GB'' have SM $Z-$fermion couplings (black and
    red), while the curves labeled ``3GB+Ferm'' allow the fermion
    couplings to vary within the $2\sigma$ region defined by
    Eq.(\ref{eq:leplim}) (blue and green).   The standard cuts given
    in Eqs.~(\ref{eq:standcuts}) and (\ref{eq:jetcuts}) are applied.}
  \label{fig:fit27}
\end{figure}

We begin by setting the fermion couplings to their SM values. In this
case the expansion to $\mathcal{O}(\Lambda^{-4})$ is the full
amplitude-squared, including both SM and SMEFT contributions. The
projections at NLO QCD for $14$~TeV are shown in Fig.~\ref{fig:fit14} and
for $27$~TeV in Fig.~\ref{fig:fit27}, in black for $\delta_{sys}=16\%$
and in red for $\delta_{sys}=4\%$, named ``3GB''. We see a significant
improvement going from $14$~TeV to $27$~TeV, while the improvement from
reducing the systematic error, $\delta_{sys} = 16\% \to 4\%$, is
marginal. Compared to Ref.~\cite{Baglio:2017bfe}, the improvement from
$8$~TeV to HL-LHC and HE-LHC is important. The coefficient $\lambda^Z$
in particular is highly constrained, $|\lambda^Z|<2\times 10^{-3}$ at
the HL-LHC and improved to $|\lambda^Z|< 6\times 10^{-4}$ at the
HE-LHC.


As demonstrated in
Refs.~\cite{Baglio:2017bfe,Grojean:2018dqj,Alves:2018nof,Liu:2018pkg,Falkowski:2016cxu},
anomalous fermion couplings can have similar effects on the $W^+W^-$
distributions as do the anomalous 3-gauge-boson couplings. At leading
order, $W^+W^-$ production proceed through two s-channel diagrams with
a photon or a $Z$ and a $t$-channel diagram.  The $s$-channel $Z$ and
$t$-channel diagrams are separately unitarity violating.  In the SM,
the violation is cancelled once the diagrams are summed and the
production rate is unitarized.  However, the $t$-channel and
$s$-channel diagrams have different dependencies on the $Z$-quark and
$W$-quark couplings.  Hence, if there are anomalous quark-gauge boson
couplings, the cancellation between the diagrams is spoiled and
perturbative unitarity is violated. Although LEP strongly constrains
these
couplings~\cite{Falkowski:2014tna,Falkowski:2017pss,Berthier:2016tkq},
these constraints are at the $Z$-pole.  At the higher energies of the
LHC, the effects of the anomalous $Z$-quark and $W$-quark couplings
grow with energy and their effect becomes important.  While the
perturbative unitarity violation is relevant for the $W^+W^-$
production, for $W^\pm$ decays into leptons the leading order is one
diagram and the process occurs at the $W$ pole.  Hence, there is no
cancellation between diagrams to guarantee unitarity conservation and
the process occurs at LEP energies. The very strong LEP constraints on
$W$-lepton couplings~\cite{Schael:2013ita} are relevant, so we set the
lepton couplings to their SM values. However we consider the effects
of anomalous 
$Z-$quark couplings, assuming flavor universality. We display in
Fig.~\ref{fig:fit14} and Fig.~\ref{fig:fit27} the constraints we obtain
on the 3-gauge-boson couplings while allowing for $Z-$quarks anomalous
couplings ranging over values that are constrained by global fits to
LEP
limits~\cite{Falkowski:2014tna,Falkowski:2017pss,Berthier:2016tkq},
\begin{eqnarray}
  \delta g_L^{Zu}&=&(-2.6\pm 1.6)\times 10^{-3},\nonumber\\
  \delta g_L^{Zd}&=&(2.3\pm 1)\times 10^{-3},\nonumber\\
  \delta g_R^{Zu}&=& (-3.6\pm 3.5)\times 10^{-3},\nonumber\\
  \delta g_R^{Zd}&=& (16.0\pm 5.2)\times 10^{-3}.
\label{eq:leplim}
\end{eqnarray}
We allow the fermion couplings to vary within the $2\sigma$ limits given
in Eq.(\ref{eq:leplim}). The curves are denoted ``3GB+Ferm'' in
Figs.~\ref{fig:fit14} and \ref{fig:fit27}, and are displayed in green
when assuming $\delta_{sys} = 4\%$ and in blue when assuming
$\delta_{sys}=16\%$. It is important to note that $\delta
g_{L/R}^{Zd}=0$ are not within the $2\sigma$ allowed range, and that
the central values for all anomalous fermions couplings are
non-zero. These observations are clearly translated into the allowed
limits for the triple gauge boson anomalous couplings in
Figs.~\ref{fig:fit14} and \ref{fig:fit27} when scanning also over the
anomalous fermions couplings in the range of Eq.~(\ref{eq:leplim}). At
$14$~TeV the only (very limited) overlap between the ``3GB'' and
``3GB+Ferm'' limits happens in the $\lambda^Z-\delta g_1^Z$ plane. The
non-zero central values of the anomalous fermion couplings interplay
with the anomalous triple gauge boson couplings, that are non-zero
 for $\delta\kappa^Z$ and $\delta g_1^Z$ as illustrated in
particular in the third plot of Fig.~\ref{fig:fit14}. This is expected
as the scan is performed relative to the SM value of the cross section: In
order to get a cross section compatible with the SM while allowing at
the same time non-zero anomalous fermion couplings, it is necessary to
allow for non-zero anomalous triple gauge boson couplings. Comparing
$14$~TeV to $27$~TeV limits we also see that there is no noticeable
improvement going from $\delta_{sys}=16\%$ down to
$\delta_{sys}=4\%$. The $2\sigma$ bounds are already saturated, and in
particular indicate non-zero values for $\delta\kappa^Z$ and $\delta
g^1_Z$ at more than $3\sigma$ when taking into account the anomalous
fermion couplings according to the fit to LEP data as given in
Eq.~(\ref{eq:leplim}). The HE-LHC will thus be able to test the
LEP fit and distinguish clearly between SM and non-SM $Z-$quark
couplings, as the curves look quite different. An important
implication of our study is that the anomalous fermion couplings have
a major result on the allowed regions and cannot be neglected.
When
comparing our results with Ref.~\cite{Grojean:2018dqj}, a rough
agreement is obtained. Our limits at the LHC are not exactly the same
and the differences can be explained by the different assumptions in
the two studies: Ref.~\cite{Grojean:2018dqj} does a background+signal
study at $13$~TeV including also contributions from the $W^\pm Z$, while
we work at $14$~TeV without taking into account the backgrounds and
focusing only on $W^+W^-$; Ref.~\cite{Grojean:2018dqj} performs a fit
on differential bins and profile over the variables not shown in their
plots, while we perform a fit on the last bin without profiling.

\begin{figure}[h!]
  {\centering
    \includegraphics[width=0.45\textwidth,clip]{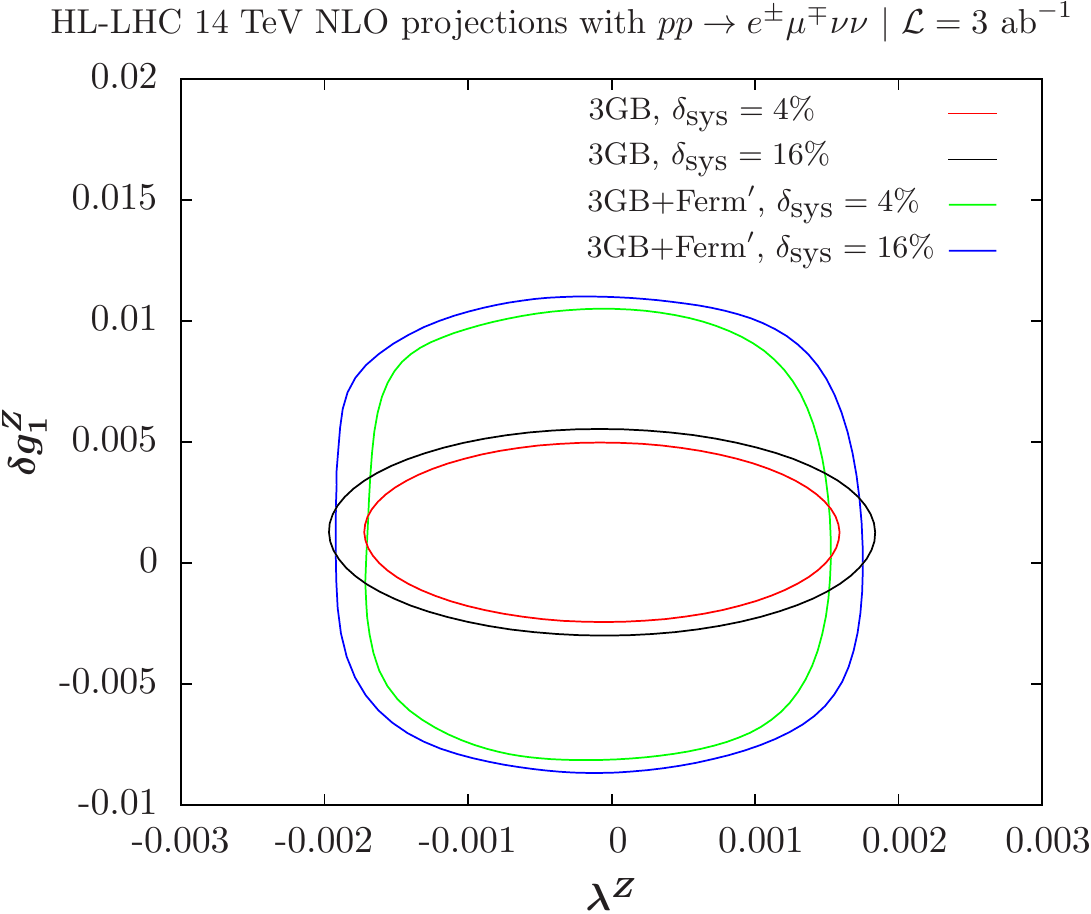}
    \includegraphics[width=0.45\textwidth,clip]{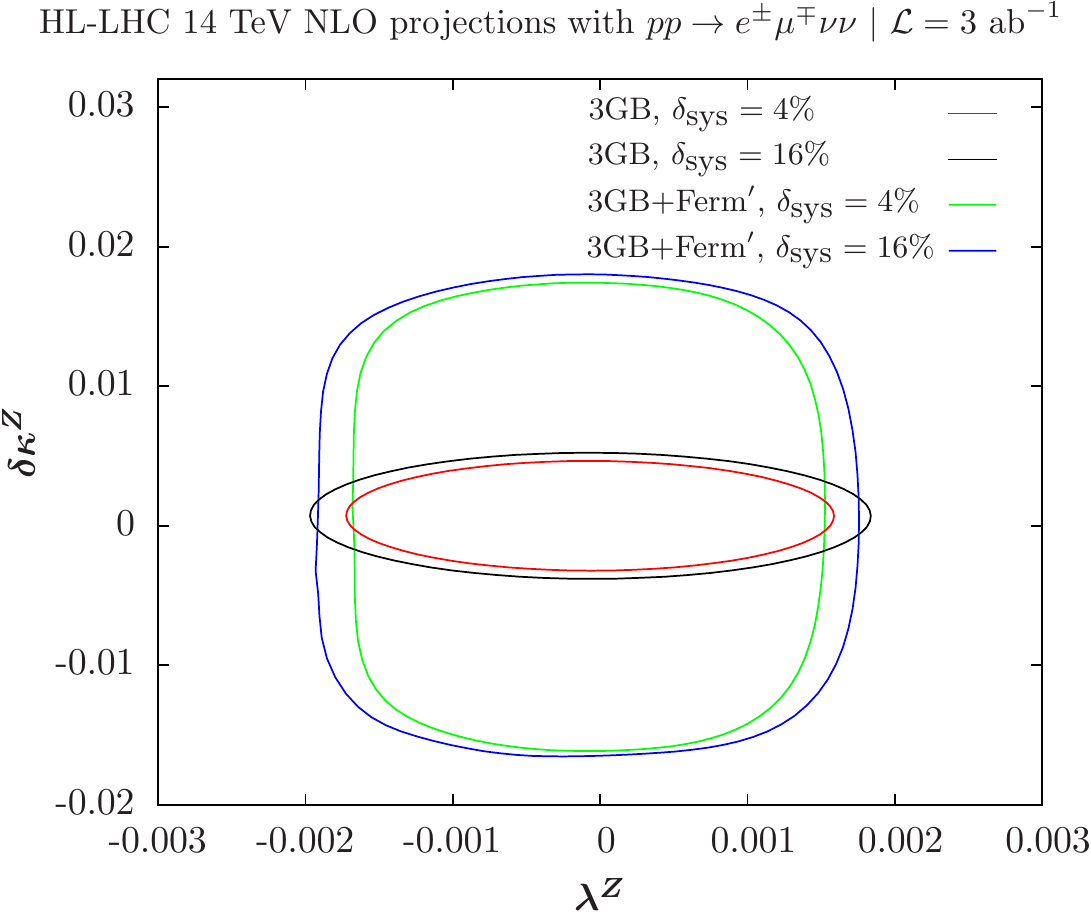}\bigskip\\
    \includegraphics[width=0.45\textwidth,clip]{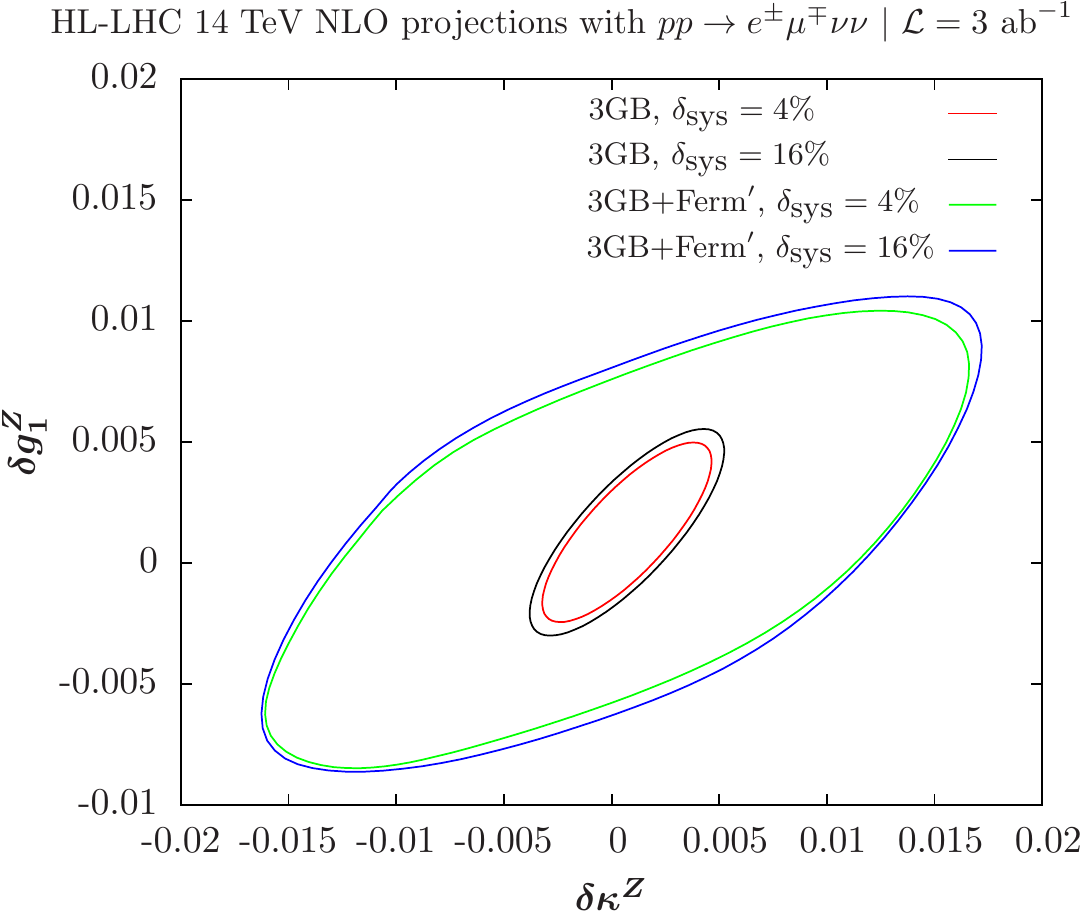}
  }
  \caption{Same as in Fig.~\ref{fig:fit14}, but with the fermion
    couplings constrained to vary around 0. The curves allowing for
    $Z$-quark coupling to vary around 0 within the $2\sigma$ limit in
    Eq.(\ref{eq:Zqlim2}) are labeled ``3GB+Ferm$^\prime$''.  The standard cuts given in Eqs.~(\ref{eq:standcuts}) and (\ref{eq:jetcuts}) are applied.}
  \label{fig:fit14_2}
\end{figure}

\begin{figure}[h!]
  {\centering
    \includegraphics[width=0.45\textwidth,clip]{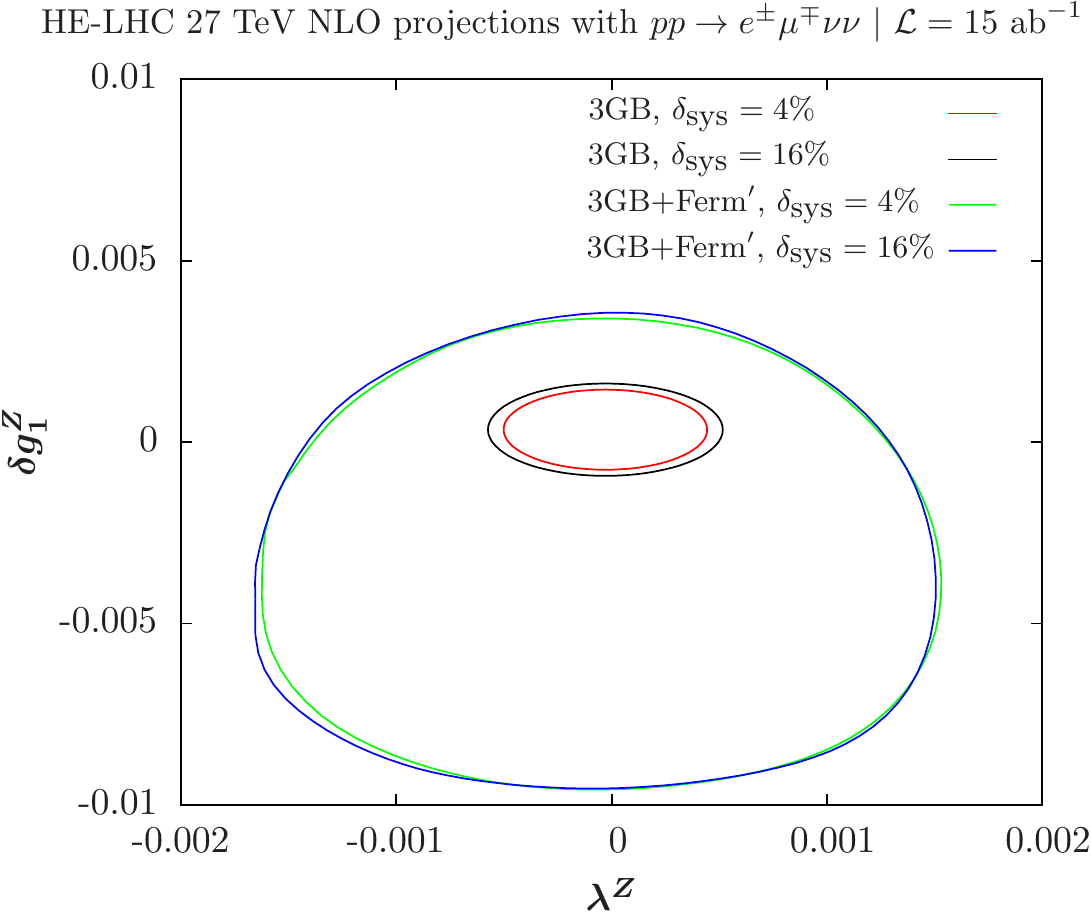}
    \includegraphics[width=0.45\textwidth,clip]{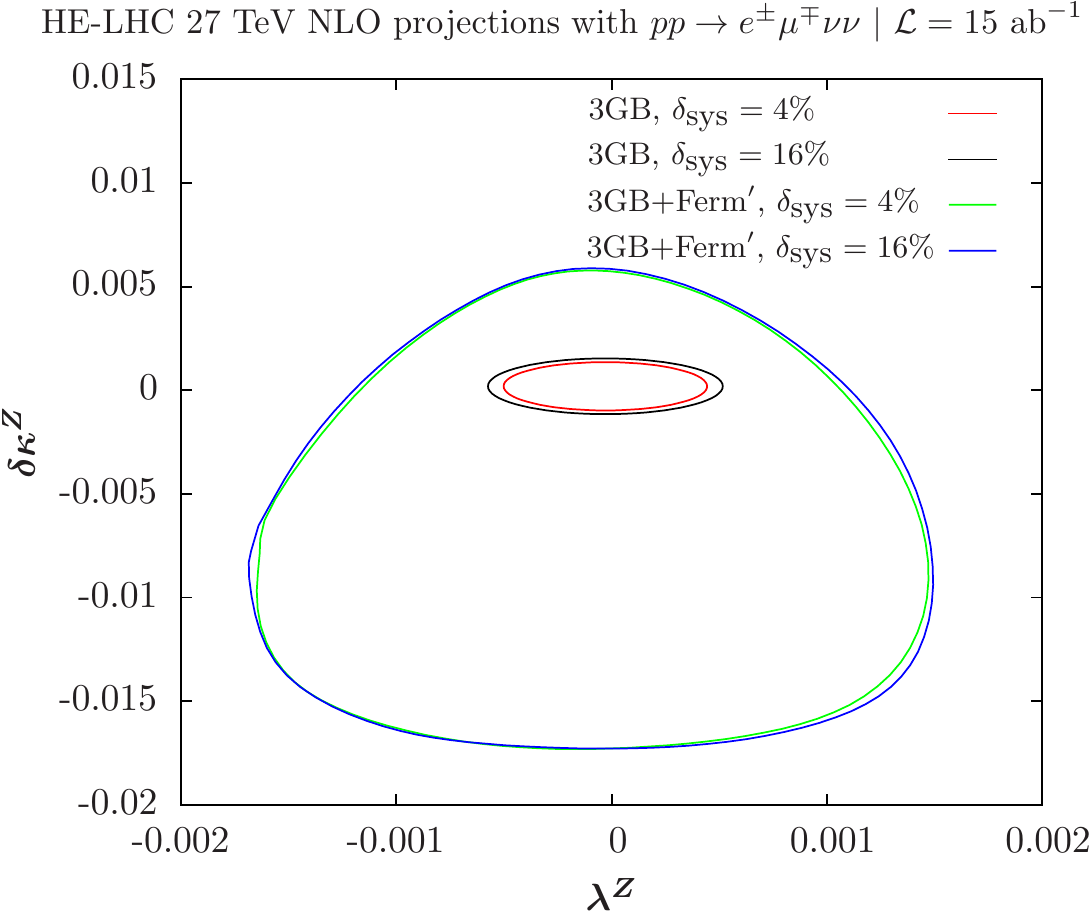}\bigskip\\    
    \includegraphics[width=0.45\textwidth,clip]{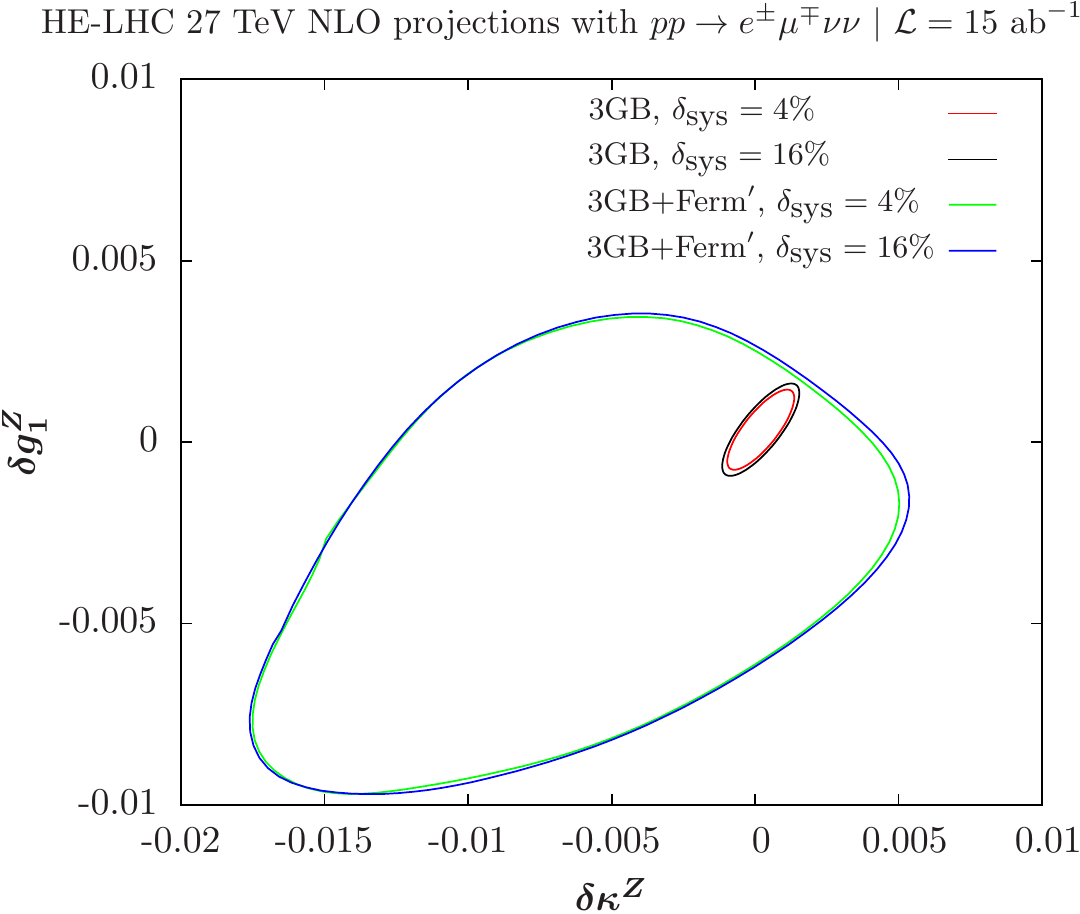}
  }
  \caption{Same as in Fig.~\ref{fig:fit27}, but with the fermion
    couplings constrained to vary around 0. The curves allowing for
    $Z$-quark coupling to vary around 0 within the $2\sigma$ limit in
    Eq.(\ref{eq:Zqlim2}) are labeled ``3GB+Ferm$^\prime$''.  The standard cuts given in Eqs.~(\ref{eq:standcuts}) and (\ref{eq:jetcuts}) are applied.}
  \label{fig:fit27_2}
\end{figure}

We have also performed a second scan where the $Z-$quark couplings are
centered around their SM value, using the same $1\sigma$ limits as in
Eq.~(\ref{eq:leplim}),
\begin{eqnarray}
  \Delta(\delta g_L^{Zu})&=&1.6\times 10^{-3},\nonumber\\
  \Delta(\delta g_L^{Zd})&=&1.0\times 10^{-3},\nonumber\\
  \Delta(\delta g_R^{Zu})&=&3.5\times 10^{-3},\nonumber\\
  \Delta(\delta g_R^{Zd})&=&5.2\times 10^{-3}.
\label{eq:Zqlim2}
\end{eqnarray}
The results are shown in Figs.~\ref{fig:fit14_2} and \ref{fig:fit27_2}
and the corresponding curves are labelled ``3GB+Ferm$^{\prime}$''. The
scan with only anomalous triple-gauge-boson couplings are given in
black and red for $\delta_{\rm sys}= 16\%$ and $4\%$ respectively, the
scan with anomalous fermion couplings centered around 0 and with
uncertainties defined according to Eq.(\ref{eq:Zqlim2}) in addition
are given in blue and green for $\delta_{\rm sys}= 16\%$ and $4\%$
respectively. The curves ``3GB+Ferm$^{\prime}$''  allow for a
central value of zero, as expected. We still see that the shapes of
the limits are no longer ellipses and that allowing for anomalous
$Z$-quark couplings worsens the limits on the anomalous
triple-gauge-boson couplings, especially at $27$~TeV. We also see in
Figs.~\ref{fig:fit27_2} the same saturation as in Figs.~\ref{fig:fit27}
when comparing the two assumptions for the systematics, the improvement from
$16\%$ down to $4\%$ is marginal. Negative values for $\delta\kappa^Z$
and $\delta g^1_Z$ are preferred over positive values. It is worth
noticing that the $27$~TeV limits for the ``3GB+Ferm$^{\prime}$'' scenario
are significantly better than those of the same scenario at $14$~TeV when comparing
Figs.~\ref{fig:fit14_2} and \ref{fig:fit27_2}, contrary to the case of
the ``3GB+Ferm'' displayed in Fig.~\ref{fig:fit14} and \ref{fig:fit27}.


\section {13 TeV distributions} 
\label{sec:13tev}

In this section, we demonstrate the use of the primitive cross
sections to compute $K$-factors and distributions in the presence of
anomalous couplings. The primitive cross sections at $13$~TeV (with
and without the basic cuts of Eqs.(\ref{eq:standcuts}) and
(\ref{eq:jetcuts})) are attached as supplemental material, including a
variety of kinematic distributions of interest. The goal is to enable
rapid scans over anomalous couplings at NLO QCD.
Our scheme is similar to the reweighting used in {\tt
  MadGraph}~\cite{Mattelaer:2016gcx}. We define  a series of
$K$-factors for the process $pp\rightarrow  W^+W^-\rightarrow \mu^\pm
e^\mp\nu{\overline{\nu}}$,
\begin{eqnarray}
K_{SM}&=&{d\sigma_{SM}^{NLO}\over d\sigma_{SM}^{LO}},\\
K_{SMEFT}^{(n)}&=&{d\sigma_{SMEFT}^{NLO}(C_1,C_2...C_m)\over d\sigma_{SMEFT}^{LO}(C_1, C_2,...C_m)},\nonumber \\
S^{(n)}_{SMEFT}&=&{d\sigma_{SMEFT}^{NLO}(C_1,C_2...C_m)\over d\sigma_{SM}^{NLO}},
\label{eq:vars}
\end{eqnarray}
where $K_{SMEFT}^{(n)}$ and $S^{(n)}_{SMEFT}$ are evaluated in the
SMEFT to ${\cal{O}}({\Lambda^{-2n}})$ and LO and NLO refer to the
order in QCD. The notation $d\sigma$ can represent either differential
or total cross sections, where the numerators and denominators must be
evaluated with identical cuts.  Applying the cuts of
Eqs.(\ref{eq:standcuts}) and (\ref{eq:jetcuts}), the SM total cross
sections are,
\begin{eqnarray}
\sigma^{LO}_{SM}&=&0.383~pb,\nonumber \\
\sigma^{NLO}_{SM}&=& 0.400~pb\, .
\label{eq:smsig}
\end{eqnarray}

\begin{figure}[tb]
{\includegraphics[width=0.45\textwidth,clip]{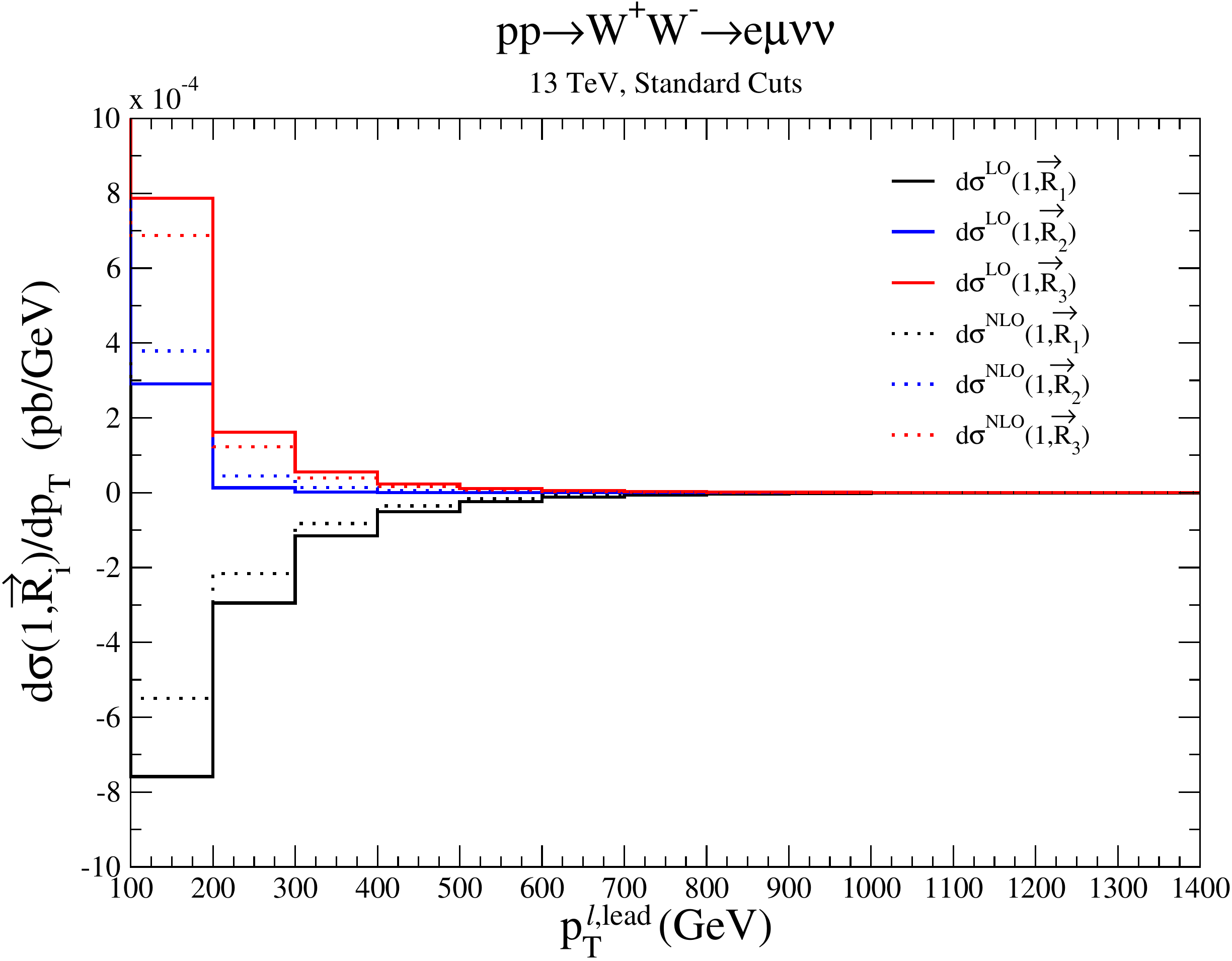}}
\caption{Primitive cross sections relevant at
  ${\cal{O}}(\Lambda^{-2})$, with $\vec{C}= (\delta g_1^Z,
  \lambda^A,\delta\kappa^Z)$.  $d\sigma(1,\vec{R}_i)$ is defined by
  Eq.(\ref{eq:prim1}). The standard cuts given in
  Eqs.~(\ref{eq:standcuts}) and (\ref{eq:jetcuts}) are applied.}
\label{fig:fitprim}
\end{figure}

\begin{figure}[tb]
{\includegraphics[width=0.45\textwidth,clip]{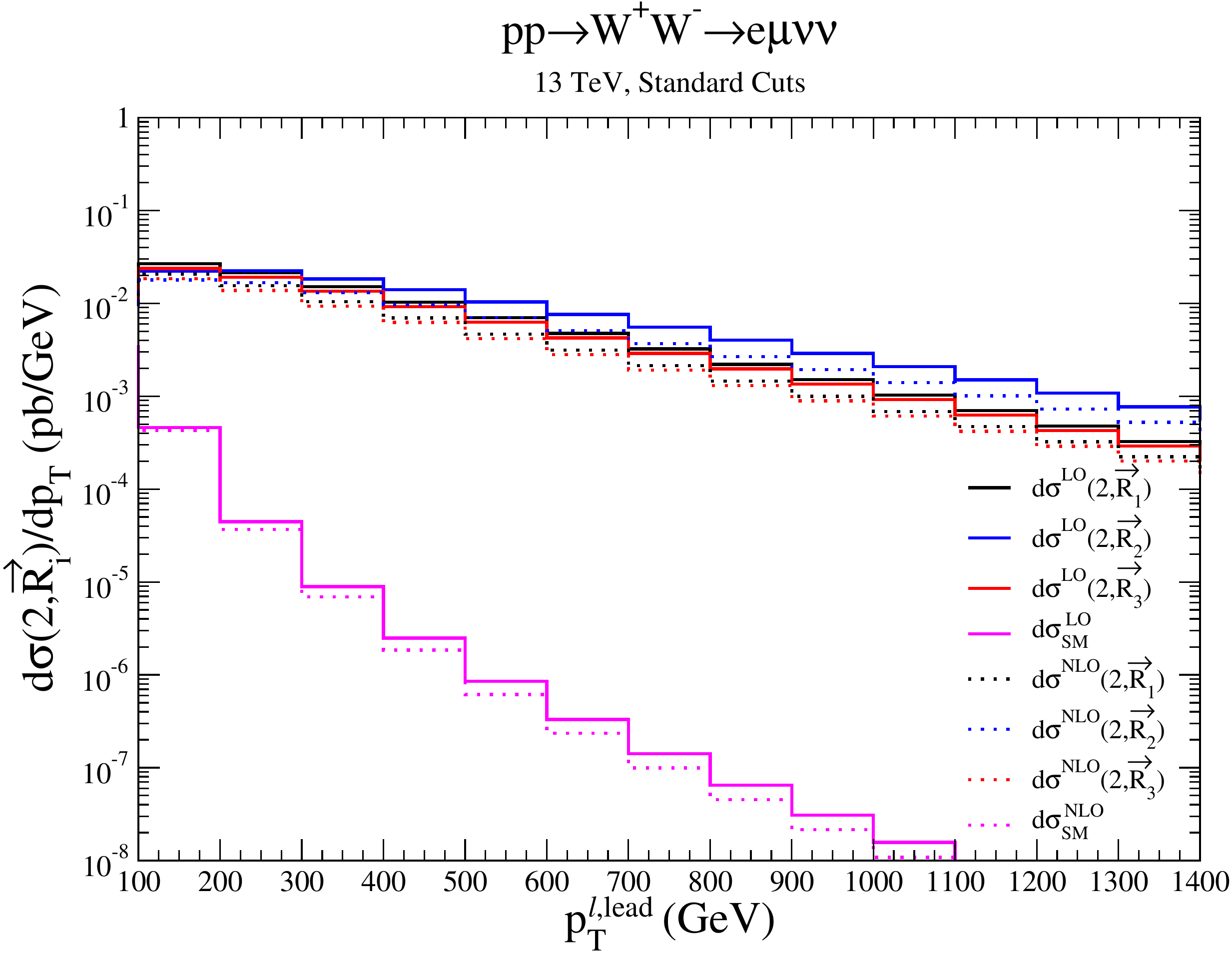}}
\hskip .5in
{\includegraphics[width=0.45\textwidth,clip]{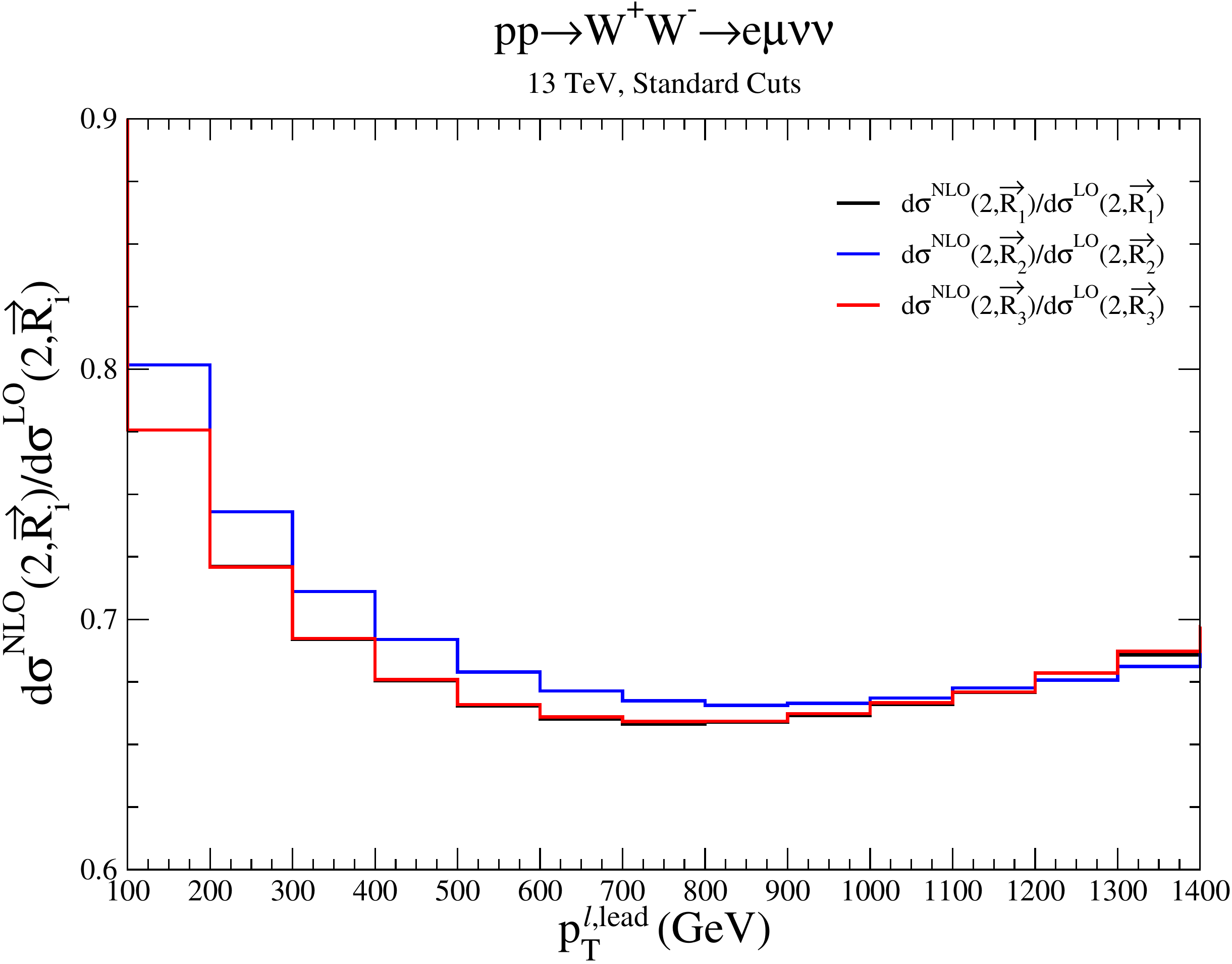}}
\caption{LHS: Primitive cross sections relevant at
  ${\cal{O}}(\Lambda^{-4})$, with $\vec{C}= (\delta g_1^Z,
  \lambda^A,\delta\kappa^Z)$.  RHS: Primitive cross sections relevant
  at ${\cal{O}}(\Lambda^{-4})$, normalized to the LO primitive cross
  sections. The black and red curves are almost
  indistinguishable. $d\sigma(2,\vec{R}_i)$ is defined in by
  Eq.(\ref{eq:prim1}). The standard cuts given in
  Eqs.~(\ref{eq:standcuts}) and (\ref{eq:jetcuts}) are applied.}
\label{fig:fitprim2}
\end{figure}

To ${\cal{O}}({ \Lambda^{-2}})$, the SMEFT $K$-factor is,
\begin{eqnarray}
K_{SMEFT}^{(1)}&=&K_{SM}\biggl[1-\Sigma_i C_i\biggl(r_{1i}^{LO}-r_{1i}^{NLO}\biggr)\biggr]\, ,
\end{eqnarray}
where we have normalized,
\begin{eqnarray}
r_{1i}^{LO}&=&{d\sigma^{LO}(1;\vec{R}_i)\over d\sigma_{SM}^{LO}},\nonumber\\
r_{1i}^{NLO}&=&{d\sigma^{NLO}(1;\vec{R}_i)\over d\sigma_{SM}^{NLO}}\, ,
\end{eqnarray}
where $d\sigma^{LO}(1;\vec{R}_i)$ and $d\sigma^{NLO}(1;\vec{R}_i)$ are
the primitive functions defined in Eq.(\ref{eq:prim1}) evaluated at LO
and NLO QCD, respectively. The primitive cross sections at ${\cal
  {O}}({\Lambda^{-2}})$ that are relevant for the case where only the
3-gauge-boson couplings are anomalous are shown in
Fig.~\ref{fig:fitprim} for the leading lepton transverse momentum,
$p_T^{\ell,lead}$. To ${\cal{O}}({\Lambda^{-4}})$, the analytic form
for the $K_{SMEFT}^{(2)}$ factor becomes rather complicated and we
evaluate it numerically using
Eq.(\ref{eq:fits}). Fig.~\ref{fig:fitprim2} shows some of the
primitive cross sections relevant at ${\cal
  {O}}({\Lambda^{-4}})$. Note that the  effects of the NLO corrections
are not the same as those in the SM, although in both cases there is a
strong dependence on $p_T^{\ell,lead}$.

A useful form for expressing the  results is,
\begin{eqnarray}
{S}_{SMEFT}^{(2)}&=& (1-\Sigma_{i=1}^{m} C_i )+\Sigma_i C_i 
\biggl[{d\sigma^{NLO}(1; {\vec{R}}_i)\over d\sigma_{SM}^{NLO}}\biggr],
\nonumber \\
{S}_{SMEFT}^{(4)}&=& (1-\Sigma_{i=1}^{m} C_i )+\Sigma_i C_i 
\biggl[{d\sigma^{NLO}(1; {\vec{R}}_i)\over d\sigma_{SM}^{NLO}}\biggr]
+\Sigma_{i=1}^m C_i^2\biggl[{d\sigma^{NLO}(2;{\vec{R}}_i)-d\sigma^{NLO}(1;{\vec{R}}_i)
\over d\sigma_{SM}^{NLO}}\biggr],
\nonumber \\
&&+\Sigma_{i>j=1}^m C_iC_j\biggl[{ d\sigma^{NLO}(2;\vec{M}_{ij})-d\sigma^{NLO}(2;{\vec{R}}_i)-d\sigma^{NLO}(2;{\vec{R}}_j)\over d\sigma_{SM}^{NLO}}+1\biggr]\, ,
\end{eqnarray}
where $d\sigma$ can be distributions or total cross sections. At
$13$~TeV with the usual cuts, see Eqs.~(\ref{eq:standcuts})  and
(\ref{eq:jetcuts}), the $S^{(4)}_{SMEFT}$ factor for the total cross
section is, 
\begin{eqnarray}
{S}_{SMEFT}^{(4)}&=& 1+  \biggl[-0.6305
C_1+        0.04881~C_2+        0.1767~C_3    +    4.541~C_4    -   0.3212~C_5\nonumber \\
&&       -4.723~ C_6    +    0.1327~  C_7 \biggr]
+19.66~C_1^2+          21.93~ C_2^2 +         21.80~C_3^2+          32.97~ C_4^2
\nonumber \\ &&
+          36.92~ C_5^2+          44.39~ C_6^2        +  25.45~ C_7^2   
+\Sigma_{i>j=1}^m C_iC_j\biggl[X_{ij}\biggr]\, .
\label{eq:keff_warsaw}
\end{eqnarray}
where the values of $X_{ij}$ are given in Table \ref{tab:xfac} and we
define the coefficients, ${\vec{C}}=(\delta g_1^Z, \lambda^Z, \delta \kappa^Z, 
\delta g_L^{Zu}, \delta g_R^{Zu},
\delta g_L^{Zd},\delta g_R^{Zd}) $.  The largest sensitivity of
$S_{SMEFT}^{(4)}$ is to the left-handed $Z$-quark couplings, as also
observed in Ref.~\cite{Grojean:2018dqj}.  Eq.(\ref{eq:keff_warsaw})
can be rescaled by the NLO SM total cross section,
Eq.(\ref{eq:smsig}), to obtain numerical values to NLO order for
arbitrary SMEFT coefficients.  Note that the numerical coefficients of
Eq.(\ref{eq:keff_warsaw}) depend on the cuts.
\begin{table}
\centering
\begin{tabular}{|l|c|c|c|c|c|c|}\hline\hline
$X_{ij}$&2&3&4&5&6&7\\ \hline
$1$ &2.211&-31.35 & -14.67&32.20&36.82 &-10.77\\
$2$ &&-4.963&-2.372&-0.1275&2.931 &0.05577\\
$3$&&&-13.49 &-48.49 &-14.66&9.676\\
$4$&&&&-0.02663&-15.00&-0.02611\\
$5$&&&&&-0.001471&-0.003671\\
$6$&&&&&&$0.0003033$ \\ \hline\hline
\end{tabular}
\caption{Coefficients defined in Eq.(\ref{eq:keff_warsaw}) with ${\vec{C}}=(\delta g_1^Z, \lambda^Z, \delta \kappa^Z, 
\delta g_L^{Zu}, \delta g_R^{Zu},
\delta g_L^{Zd},\delta g_R^{Zd} )$  .  The standard cuts given in Eqs.~(\ref{eq:standcuts}) and (\ref{eq:jetcuts}) are applied.}
\label{tab:xfac}
\end{table}

For comparison, we present $S^{(4)}_{SMEFT}$ in the HISZ basis of
Eq.(\ref{eq:hiszbas}) using the primitive cross sections discussed
above. Using the results of the previous section, $C_i=\Sigma_j
\alpha_{ij}C_j^\prime$, with, 
\begin{equation}
C_i=\left(\begin{matrix}\delta g_1^Z\\ \delta \kappa_Z\\
\lambda_Z\end{matrix}\right), \qquad   C_i^\prime =\left(\begin{matrix}{\hat{f_W}}=f_W
\biggl({M_Z^2\over  \Lambda^2} \biggr)
\\ {\hat{f_B}}=f_B\biggl({M_Z^2\over  \Lambda^2} \biggr)
\\
{\hat{f}}_{WWW}=f_{WWW}\biggl({M_Z^2\over  \Lambda^2} \biggr)\end{matrix}\right),\qquad 
\alpha=\left(\begin{matrix}
{1\over 2}& 0 & 0\\
{c_W^2\over 2}  & -{s_W^2\over 2} & 0\\
0 & 0& {3g^2c_W^2\over 4}\end{matrix}\right)\, ,
\end{equation}
and $C_i=C_i^\prime,~i=4-7$.
The $S^{(4)}_{SMEFT}$ factor in a transformed basis is,
\begin{eqnarray}
S_{SMEFT}^{(4)}&=& (1-\Sigma_{i,k=1}^{m} C_i^\prime \alpha_{ki} )+\Sigma_{i,k=1}^m C_i^\prime \alpha_{ki}
\biggl[{d\sigma^{NLO}(1; {\vec{R}}_k)\over d\sigma_{SM}^{NLO}}\biggr]\nonumber \\
&&
+\Sigma_{i=1}^m C_i^{\prime~2}\biggl\{
\Sigma_{k=1}^m \alpha_{ki}^2\biggl[{d\sigma^{NLO}(2;\vec{R}_k)-d\sigma^{NLO}(1;\vec{R}_k)\over d\sigma_{SM}^{NLO}}\biggr]
\nonumber \\ 
&&+\Sigma_{l=1}^m\Sigma_{k=l+1}^m\alpha_{ki}\alpha_{li}
\biggl[{d\sigma^{NLO}(2,\vec{M}_{kl})-
d\sigma^{NLO}(2;{\vec{R}_k})-d\sigma^{NLO}(2;{\vec{R}_l})\over d\sigma_{SM}^{NLO}} +1\biggr]\biggr\}
\nonumber \\
&&+\Sigma_{j=1}^m\Sigma_{i=j+1}^m  C_i^\prime C_j^\prime 
\biggl\{ \Sigma_{k=1}^m 2\alpha_{ki}\alpha_{kj}
\biggl[{d\sigma^{NLO}(2;\vec{R}_k)-d\sigma^{NLO}(1;\vec{R}_k)\over d\sigma_{SM}^{NLO}}\biggr]\nonumber \\
&&+\Sigma_{l=1}^m\Sigma_{k=l+1}^m\biggl(\alpha_{kj}\alpha_{li}+\alpha_{ki}\alpha_{lj}\biggr)
\nonumber \\ && \cdot
\biggl[{d\sigma^{NLO}(2;\vec{M}_{kl})-d\sigma^{NLO}(2;\vec{R}_k)-d\sigma^{NLO}(2;\vec{R}_l)\over d\sigma_{SM}^{NLO}}+1\biggr]
\biggr\}\, .
\end{eqnarray}
For $i,j=4-7$, $\alpha_{ij}=\delta_{ij}$. The primitive cross sections
$d\sigma^{NLO}(n,\vec{R}_i)$ and $d\sigma^{NLO}(n,\vec{M}_{ij})$ are
defined in Eqs.~(\ref{eq:prim1},\ref{eq:prim2}) and are evaluated in
the original anomalous couplings basis.

For the total cross section  in the HISZ basis applying our basic cuts
in Eqs.~(\ref{eq:standcuts}) and (\ref{eq:jetcuts}) with
${\vec{C}^\prime} = ({\hat{f_W}},  {\hat{f_B}},  {\hat{f}}_{WWW}, 
\delta g_L^{Zu}, \delta g_R^{Zu},
\delta g_L^{Zd},\delta g_R^{Zd} )$  and $\Lambda=1$~TeV,
\begin{eqnarray}
S^{(4)}_{SMEFT,~HISZ}&=& 1+   \biggl[
- 0.4059 C_1^\prime  
   -  1.115C_2^\prime
 -0.7189 C_3^\prime
     \nonumber \\ &&
     +      
    4.541    C_4^\prime -      0.3211  C_5^\prime       
    -4.723   C_6^\prime +      0.1327   C_7^\prime\biggr]
     \nonumber \\    
&&
+  8.687C^{\prime~2}_1+         
 .2639C^{\prime~2} _2 +         
1.2C^{\prime~2}_3+          32.97 C^{\prime~2}_4
\nonumber \\ &&
+          36.92 C^{\prime~2} _5+          44.39 C^{\prime~2}_6        +
  25.45 C^{\prime~2} _7   
+\Sigma_{i>j=1}^m C_i^\prime C_j^\prime \biggl[{\hat{X}}_{ij}\biggr]\, .
\label{eq:keff_hisz}
\end{eqnarray}
where the values of ${\hat{X}}_{ij}$ are given in Table \ref{tab:xfac_hisz}. 
\begin{table}
\centering
\begin{tabular}{|l|c|c|c|c|c|c|}\hline\hline
$\hat{X}_{ij}$&2&3&4&5&6&7\\ \hline
$1$ &-1.447&-3.917 & -14.67&32.20&36.82 &-10.77\\
$2$ &&1.178&-2.372&-0.1275&2.931 &0.05577\\
$3$&&&-13.49 &-48.49 &-14.66&9.676\\
$4$&&&&-0.02663&-15.00&-0.02611\\
$5$&&&&&-0.001471&-0.003671\\
$6$&&&&&&$0.0003033$ \\ \hline\hline
\end{tabular}
\caption{HISZ basis coefficients defined in Eq.~(\ref{eq:keff_hisz})
  with ${\vec{C}^\prime}=({\hat{f}}_W, {\hat{f}}_B, {\hat{f}}_{WWW}, 
\delta g_L^{Zu}, \delta g_R^{Zu},
\delta g_L^{Zd},\delta g_R^{Zd}) $  and $\Lambda=1$~TeV.  The standard
cuts given in Eqs.~(\ref{eq:standcuts}) and (\ref{eq:jetcuts}) are
applied.}
\label{tab:xfac_hisz}
\end{table}

\begin{figure}[h]
{\includegraphics[width=0.45\textwidth,clip]{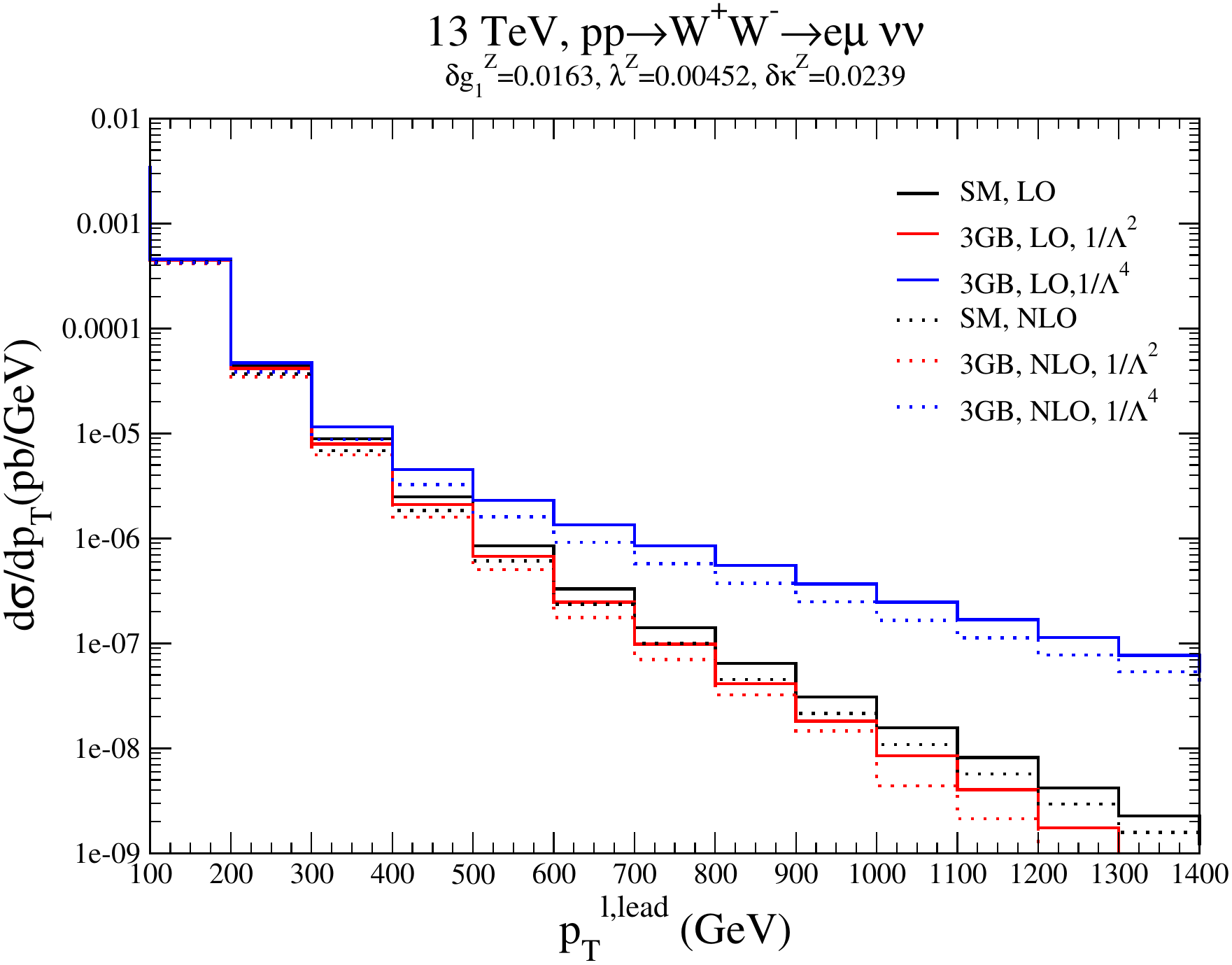}}
\hskip .5in
{\includegraphics[width=0.45\textwidth,clip]{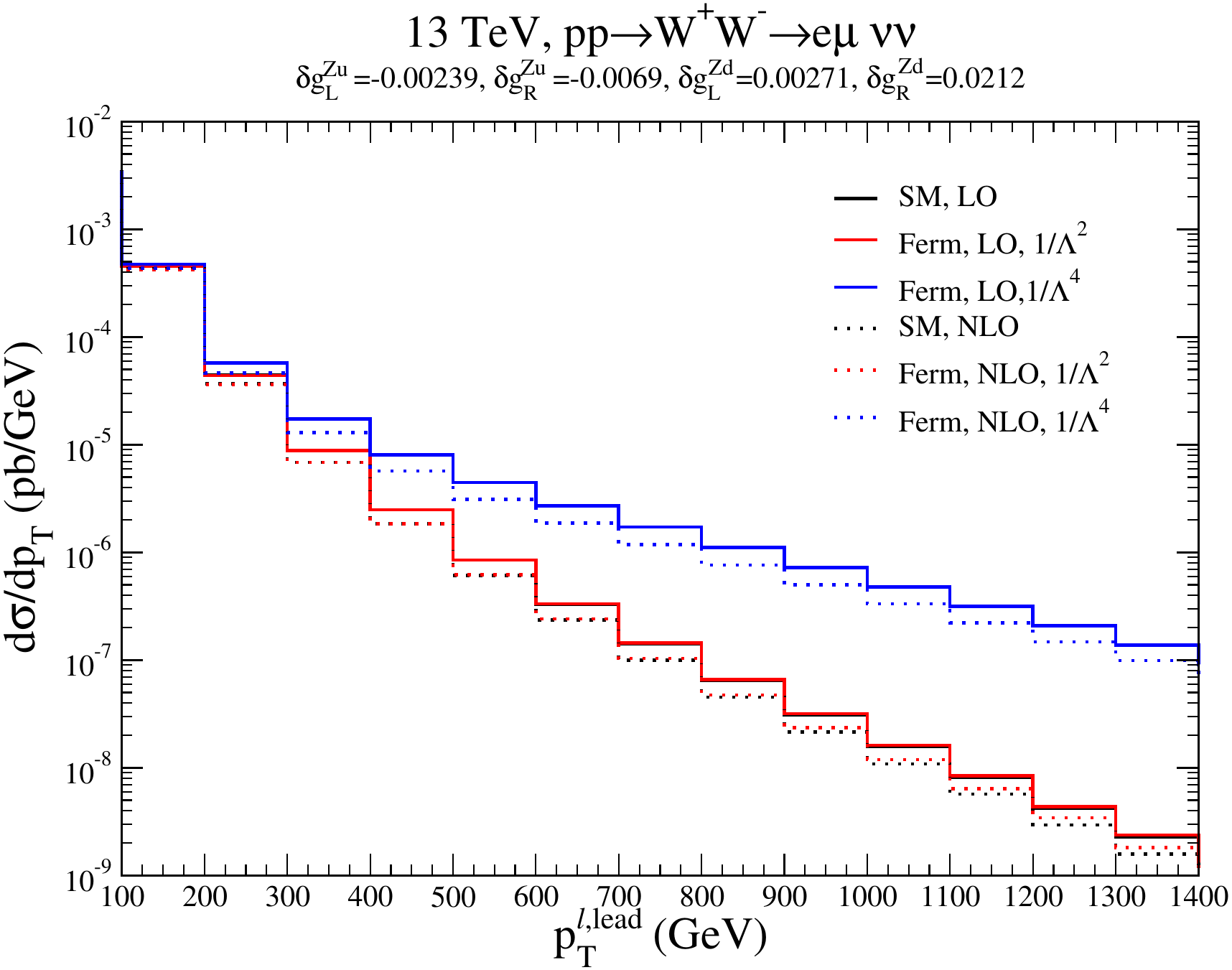}}
\caption{Distribution of the leading charged lepton $p_T$ in a
  scenario with only anomalous 3-gauge-boson couplings (LHS) and only
  anomalous $Z-$ fermion couplings (RHS) to ${\cal {O}}(\Lambda^{-2})$
  and ${\cal {O}}(\Lambda^{-4})$ at LO and NLO QCD. The SM, LO and
  Ferm, LO, $1/\Lambda^2$ curves on the RHS are indistinguishable. The
  standard cuts given in Eqs.~(\ref{eq:standcuts}) and
  (\ref{eq:jetcuts}) are applied.}
\label{fig:ptdist_eft}
\end{figure}

The primitive cross sections can also be used to study distributions with
arbitrary SMEFT coefficients and we show some sample results for the
transverse momentum of the leading charged lepton, $p_T^{\ell,lead}$, and for the
invariant mass distribution of the charged leptons, $m_{\ell\ell}$.  In
Fig.~\ref{fig:ptdist_eft}, we show the distribution of the leading
lepton $p_T$ in a scenario with only anomalous 3-gauge-boson couplings
(LHS) and with only anomalous $Z-$ quark couplings (RHS).  The values
of the coefficients were chosen to be allowed by experimental limits
from $W^+W^-$ pair production~\cite{Aad:2016wpd,Khachatryan:2015sga}
and from fits to LEP data~\cite{Schael:2013ita}, and to give similar
$p_T$ distributions. It is apparent that the anomalous
3-gauge-boson-only and anomalous-$Z-$fermion-only scenarios cannot be
distinguished by studying the $p_T$ distributions alone and that the
dominant effects come from the $\Lambda^{-4}$ contributions. The NLO
effects decrease the rate at high $p_T$ for the anomalous
3-gauge-boson couplings and increase it for anomalous $Z-$ fermion
couplings.

\begin{figure}[h]
{\includegraphics[width=0.45\textwidth,clip]{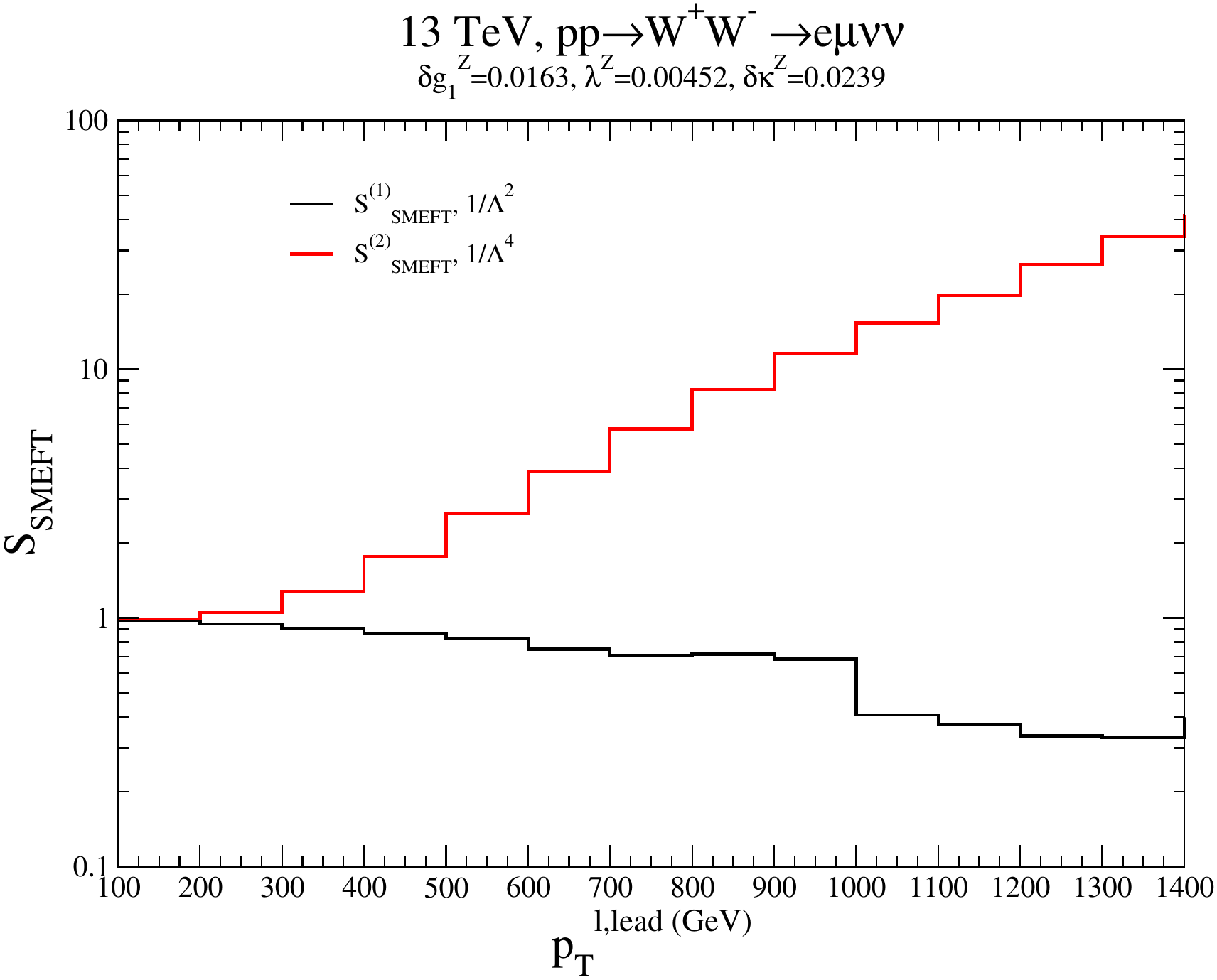}}
\hskip .5 in
{\includegraphics[width=0.45\textwidth,clip]{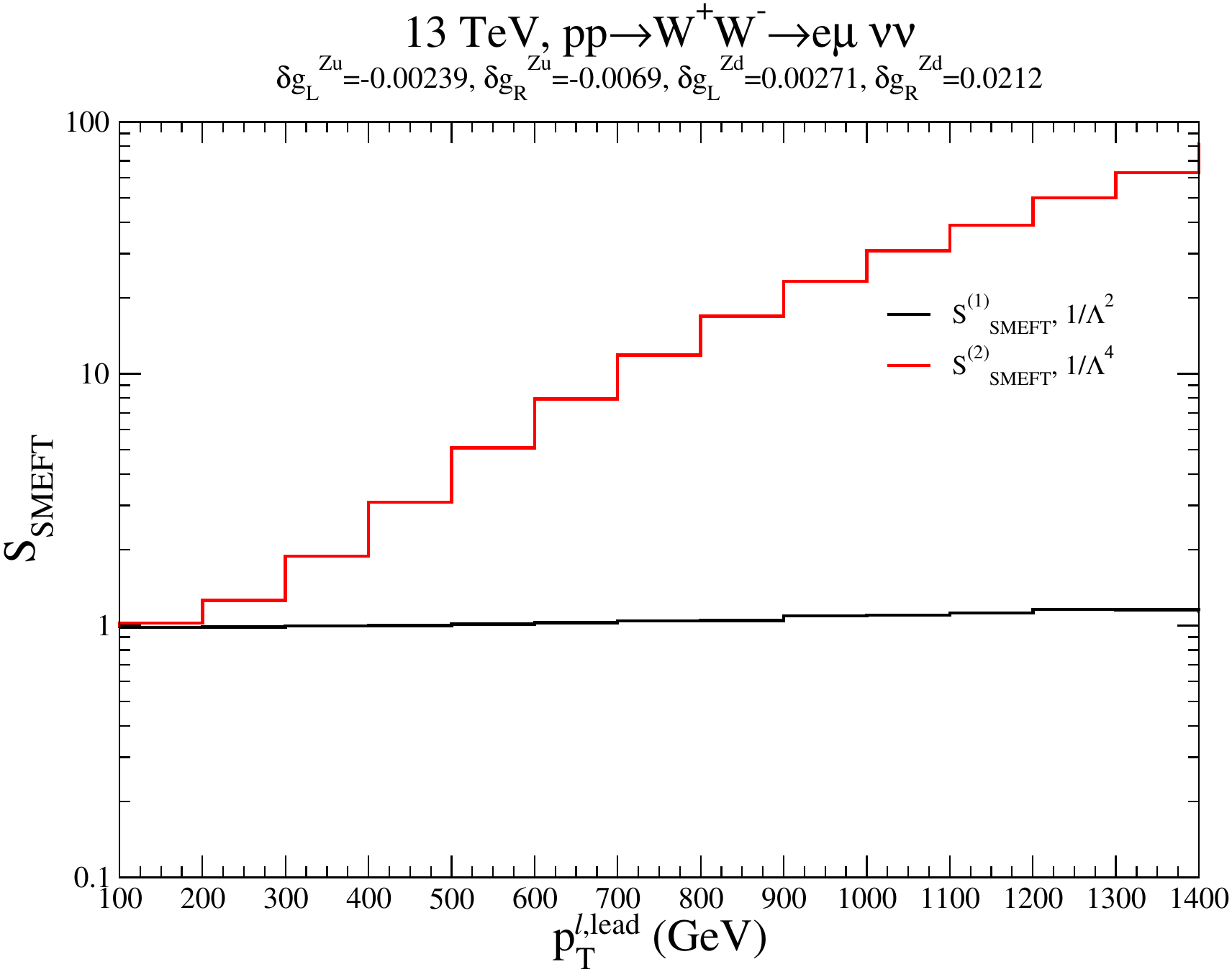}}
\caption{Comparison of the $S$  factors of Eq.(\ref{eq:vars}) for the
  SM and SMEFT in the anomalous 3-gauge-boson-only scenario (LHS) and
  anomalous-fermion-only scenario (RHS) at ${\cal{O}}({\Lambda^{-2}})$
  and ${\cal{O}}({\Lambda^{-4}})$ .  The standard cuts given in
  Eqs.~(\ref{eq:standcuts}) and (\ref{eq:jetcuts}) are applied.}
\label{fig:ptk}
\end{figure}

\begin{figure}[h]
{\includegraphics[width=0.45\textwidth,clip]{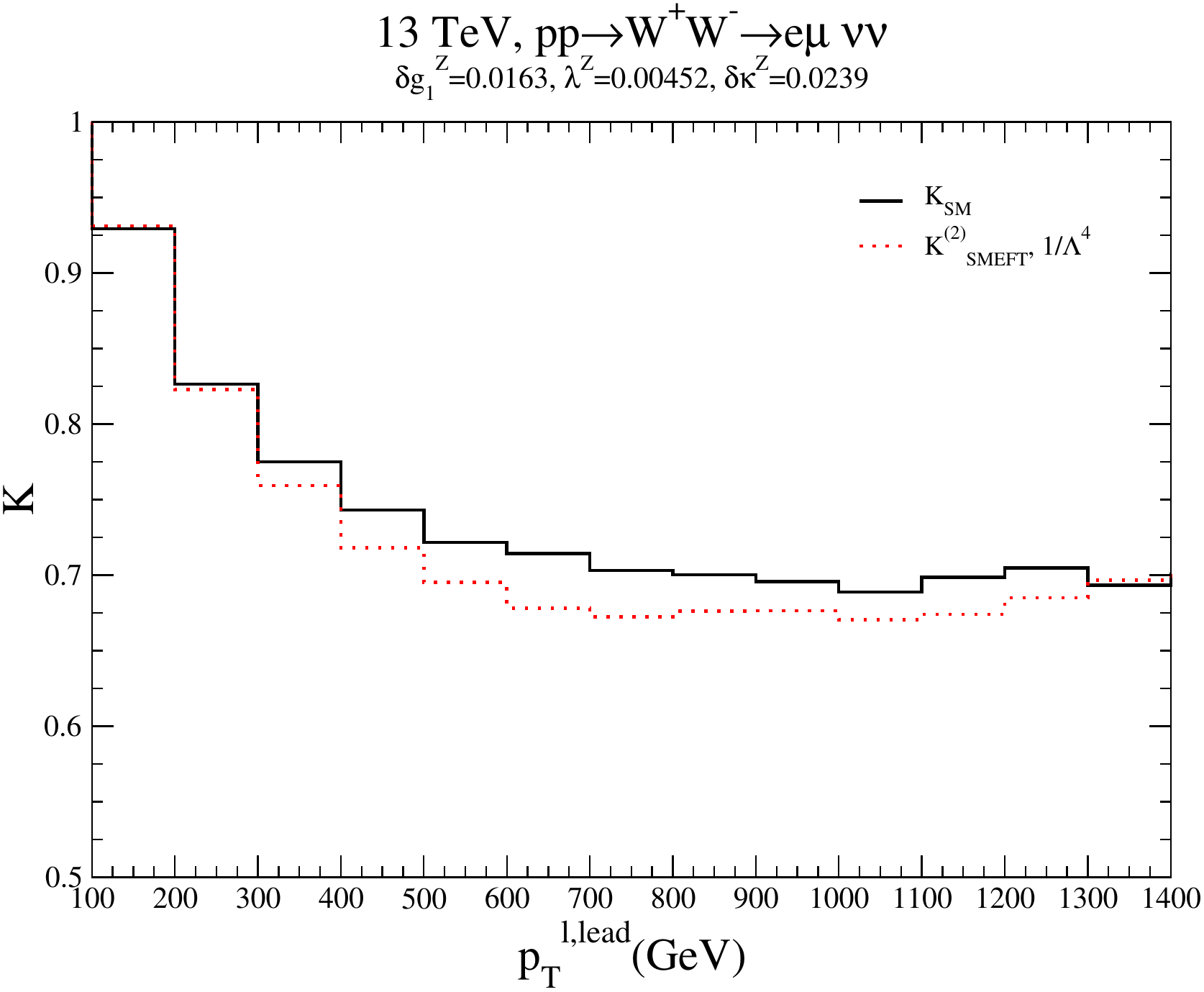}}
\hskip .5in
{\includegraphics[width=0.45\textwidth,clip]{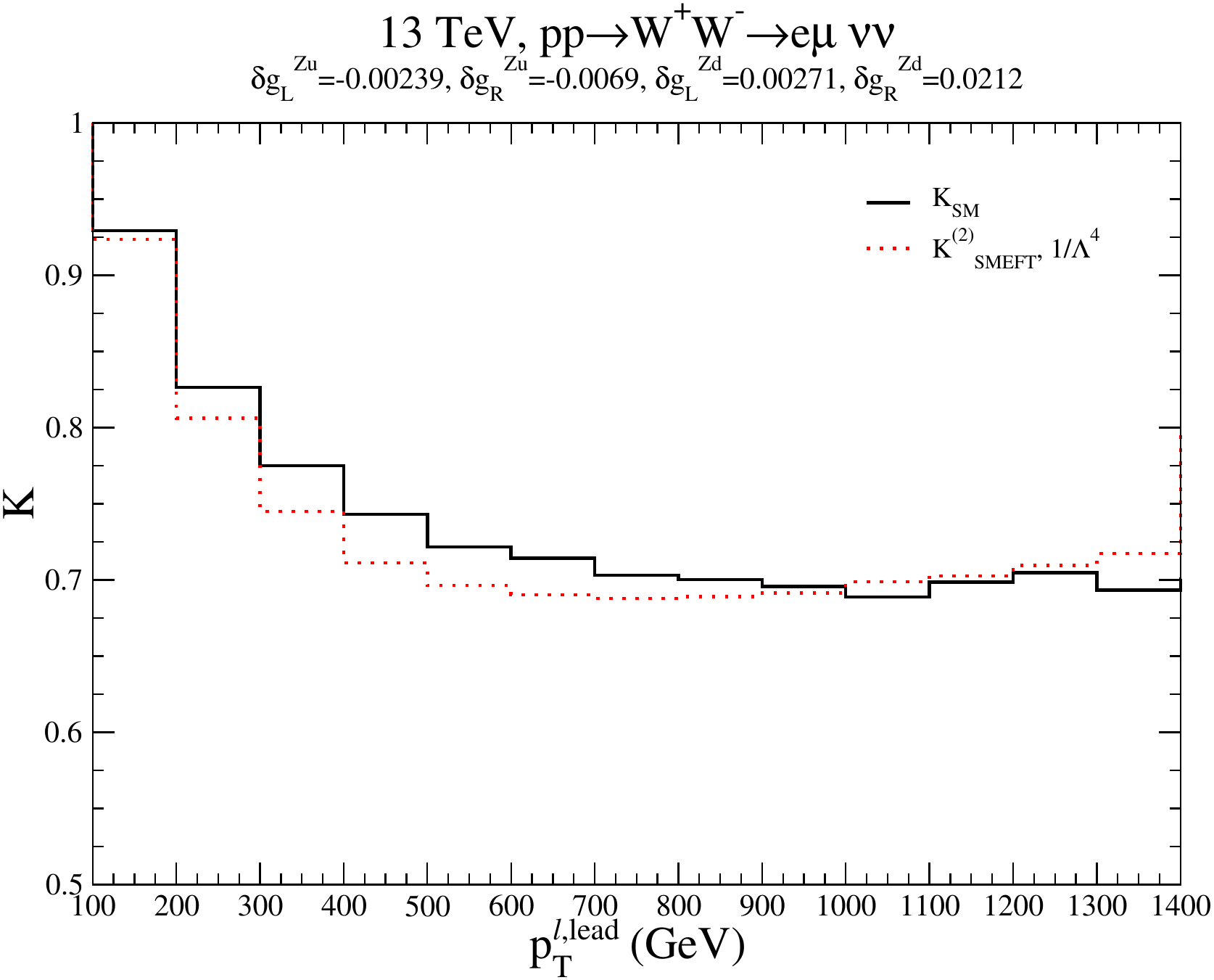}}
\caption{Comparison of the $K$-factors of Eq.(\ref{eq:vars}) for the
  SM and SMEFT in the anomalous 3-gauge-boson-only scenario (LHS) and
  anomalous-fermion-only scenario (RHS).  The standard cuts given in
  Eqs.~(\ref{eq:standcuts}) and (\ref{eq:jetcuts}) are applied.}
\label{fig:ptk_new}
\end{figure}

Figs.~\ref{fig:ptk} and \ref{fig:ptk_new} we show the variables of
Eq.(\ref{eq:vars}).  The $S^{(n)}_{SMEFT}$ variable compares 
the SMEFT distributions with those of the SM, demonstrating the
dominance of the $\Lambda^{-4}$ terms at large $p_T$. SM and SMEFT
$K$-factors are shown in Fig.~\ref{fig:ptk_new} and it is clear that
the SM and SMEFT scale similarly with small variations. Similarly, the
K-factors for the anomalous-fermion-only scenario (RHS or
Fig.~\ref{fig:ptk_new}) and the anomalous-3GB-only scenario (LHS of
Fig.~\ref{fig:ptk_new}) are also similar with small variations.  We
note that this is for specific choices of the anomalous couplings and
for different choices the $K$-factors have to be checked using the
primitive cross sections.

\begin{figure}[h]
{\includegraphics[width=0.45\textwidth,clip]{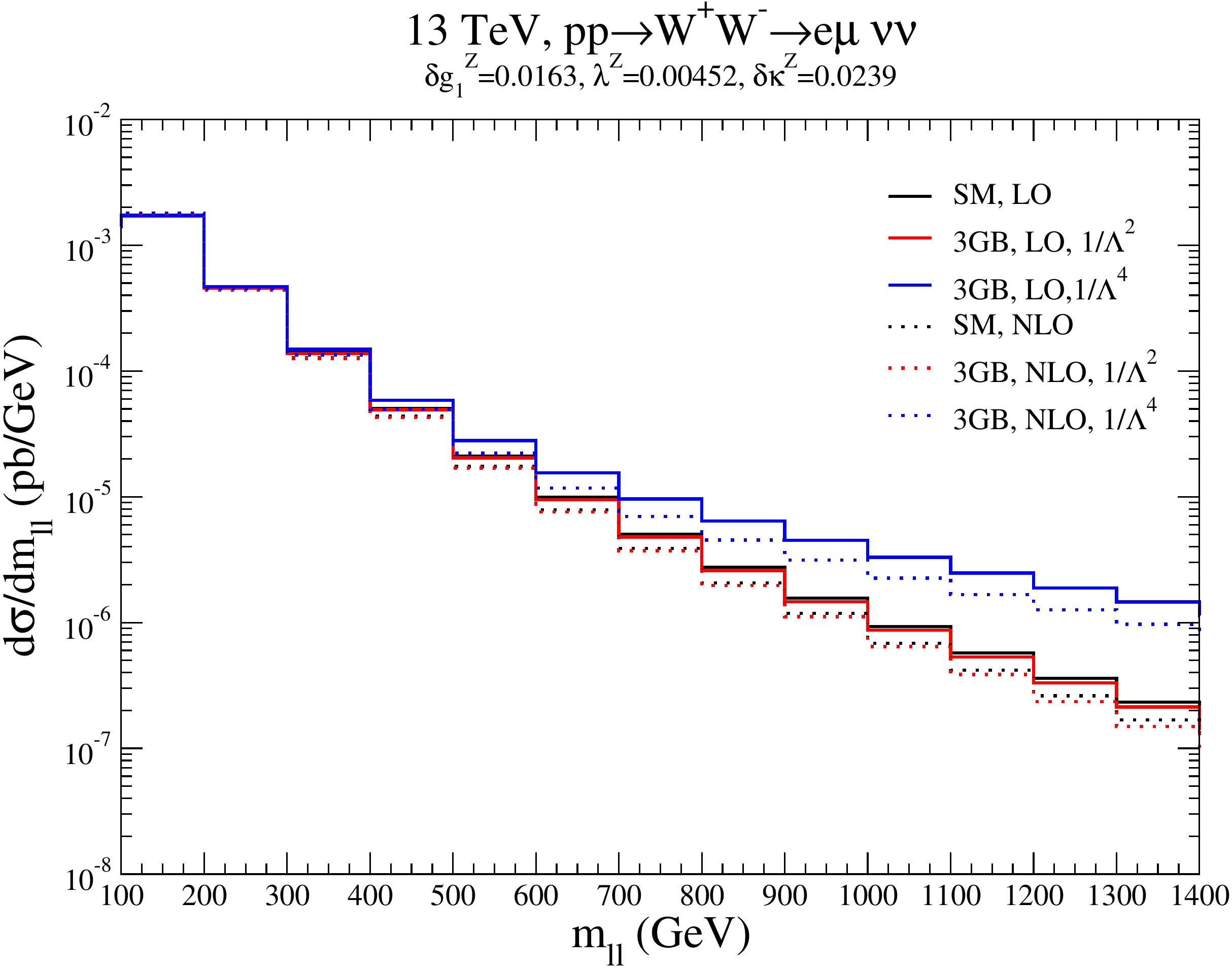}}
\hskip .5in
{\includegraphics[width=0.45\textwidth,clip]{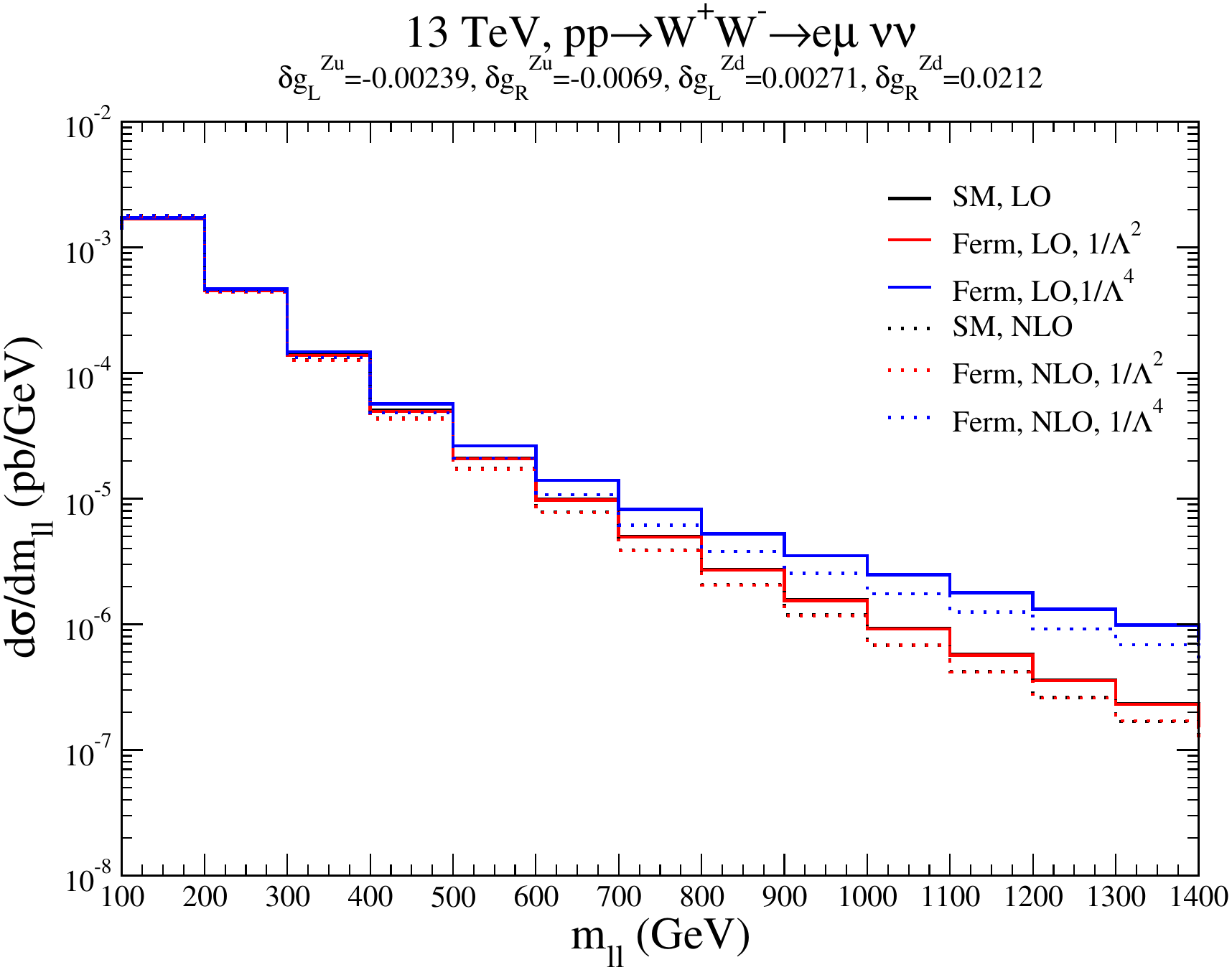}}
\caption{Invariant mass distribution  in a scenario with only
  anomalous 3-gauge-boson couplings (LHS) and only anomalous fermion
  couplings (RHS) to ${\cal {O}}(\Lambda^{-2})$ and ${\cal
    {O}}(\Lambda^{-4})$  at LO and NLO QCD.  The standard cuts given
  in Eqs.~(\ref{eq:standcuts}) and (\ref{eq:jetcuts}) are applied.}
\label{fig:mll}
\end{figure}

\begin{figure}[h]
{\includegraphics[width=0.45\textwidth,clip]{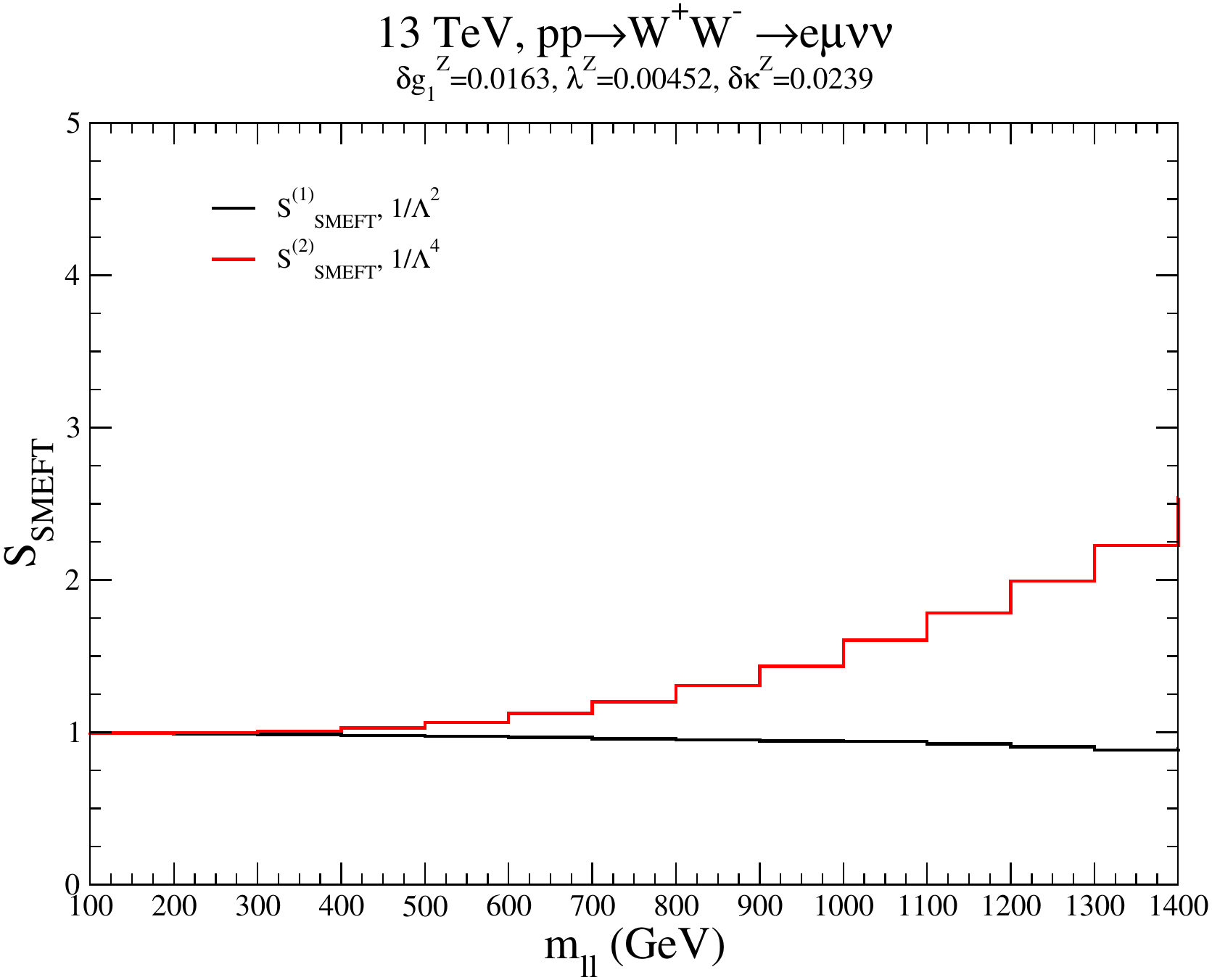}}
\hskip .5in
{\includegraphics[width=0.45\textwidth,clip]{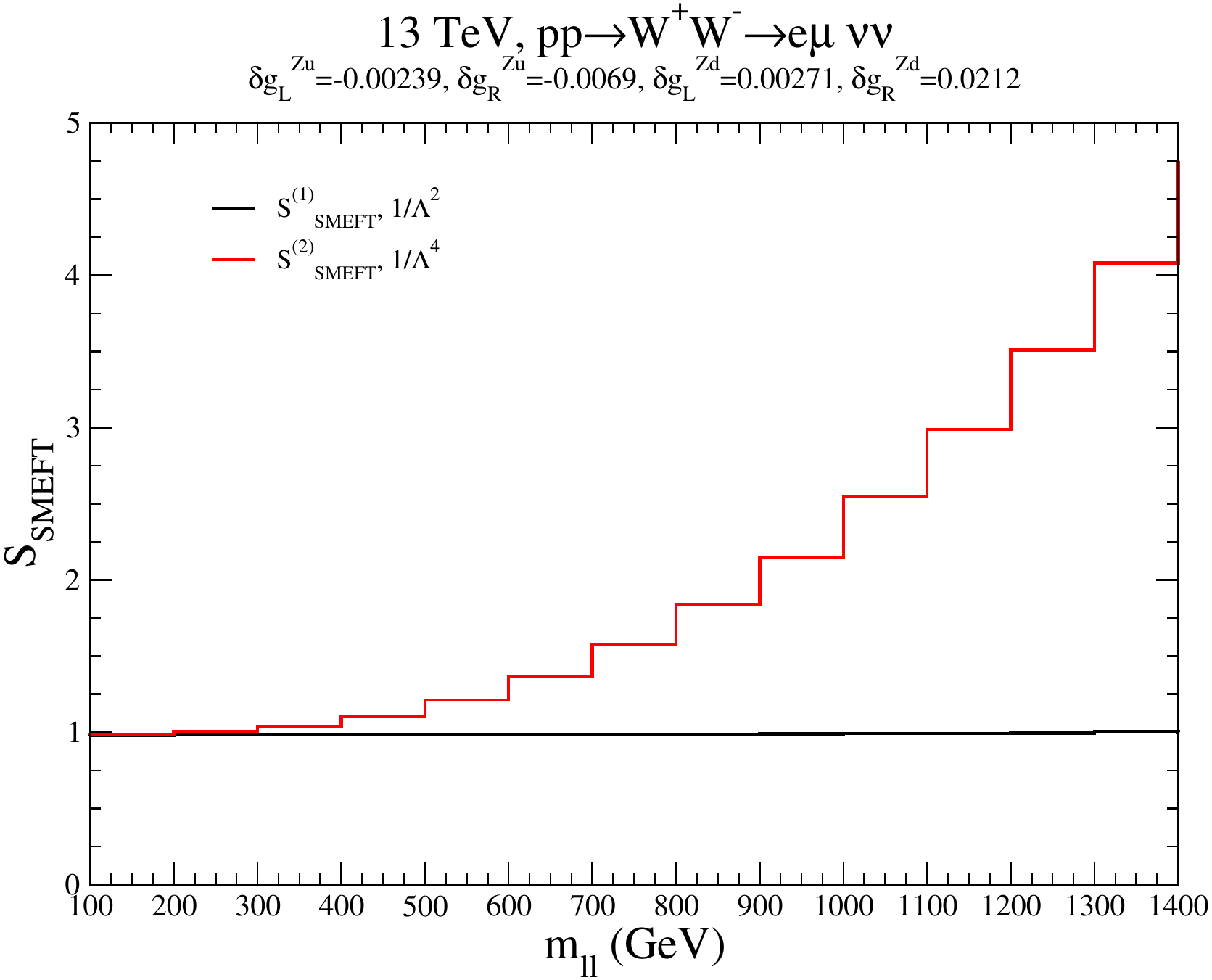}}
\caption{Comparison of the $S$  factors of Eq.(\ref{eq:vars}) for the
  SM and SMEFT in  the anomalous 3-gauge-boson-only scenario (LHS) and
  anomalous-fermion-only scenario (RHS).  The standard cuts given in
  Eqs.~(\ref{eq:standcuts}) and (\ref{eq:jetcuts}) are applied.}
\label{fig:k_mll}
\end{figure}

\begin{figure}[h]
{\includegraphics[width=0.45\textwidth,clip]{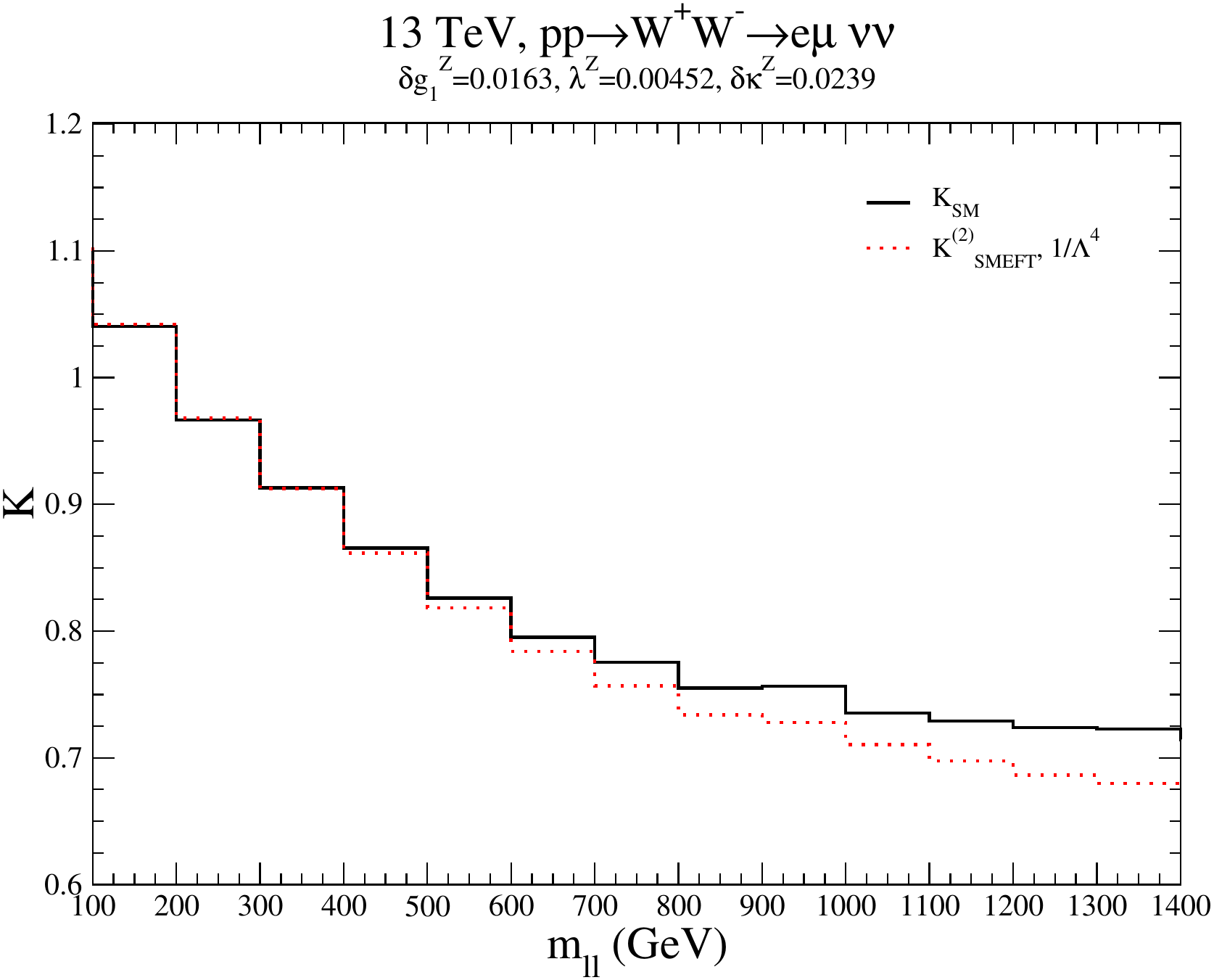}}
\hskip .5in
{\includegraphics[width=0.45\textwidth,clip]{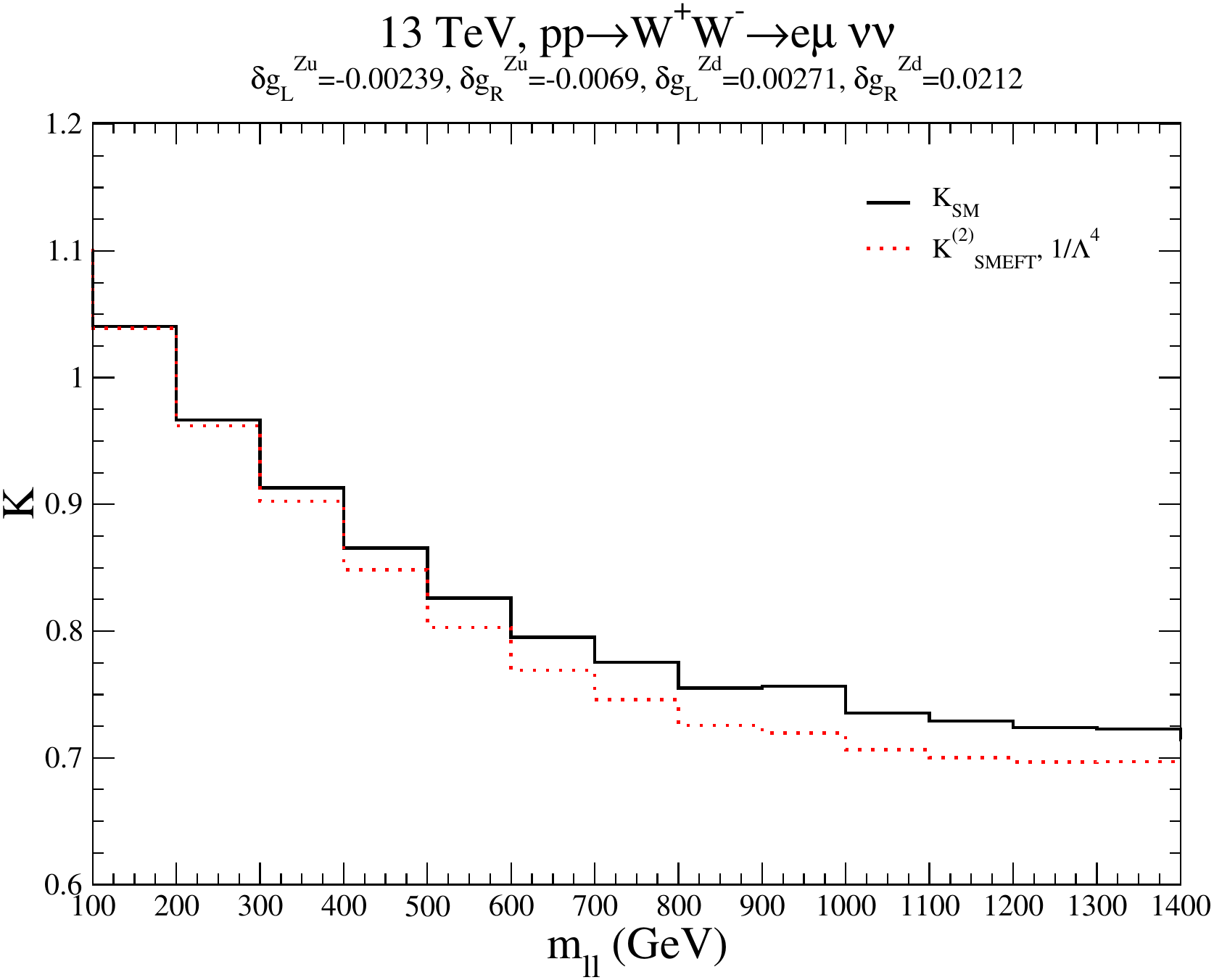}}
\caption{Comparison of the $K$-factors of Eq.(\ref{eq:vars}) for the
  SM and SMEFT in the anomalous 3-gauge-boson-only scenario (LHS) and
  anomalous-fermion-only scenario (RHS). The standard cuts given in
  Eqs.~(\ref{eq:standcuts}) and (\ref{eq:jetcuts}) are applied.}
\label{fig:k_mll_new}
\end{figure}

In Fig.~\ref{fig:mll}, we show the invariant mass distribution of the
charged leptons in a scenario with only anomalous 3-gauge-boson
couplings (LHS) and with only anomalous $Z-$ quark couplings (RHS).
As is the case with the $p_T$ distributions, the dominant effect
arises from the $\Lambda^{-4}$ contributions.
Figs.~\ref{fig:k_mll} and \ref{fig:k_mll_new} show the variables of
Eq.(\ref{eq:vars}). The $S^{(n)}_{SMEFT}$ variable compares the SMEFT
distributions with those of the SM, again showing the large effects
from  the ${\cal{O}}(\Lambda^{-4})$ terms. SM and SMEFT $K$-factors
are shown in Fig.~\ref{fig:k_mll_new} and are quite similar to each
other. 

Ref.~\cite{Chiesa:2018lcs} computes the NLO QCD (along with the NLO
EW) corrections to gauge boson pair production when the fermion
couplings take their SM values.  Our results are qualitatively similar
to theirs for the scenario with only anomalous 3-gauge-boson
couplings.

\section{Conclusions} 
We have extended our previous NLO calculation of the contribution of
anomalous couplings to $pp\rightarrow W^+W^-$ to include the leptonic
decays, $pp\rightarrow W^+W^-\rightarrow \mu^\pm e^\mp \nu
{\overline{\nu}}$ and implemented the results in the {\tt POWHEG-BOX}.
The primitive cross sections at $13$~TeV for a variety of observables are
posted at
\url{https://quark.phy.bnl.gov/Digital_Data_Archive/dawson/ww_18}.
The most important implication of our results is the interplay between
anomalous $Z$-quark couplings and anomalous 3-gauge-boson couplings
requiring global fits to both fermion and gauge couplings in order to
obtain reliable results.  Our NLO results at $13$~TeV are presented in
terms of primitive cross sections allowing for rapid scans over
anomalous couplings in an arbitrary basis.  

\label{sec:conc}

\begin{acknowledgments}
We thank Giulia Zanderighi and Paolo Nason for discussions about the
private updated version of their code. We also thank Barbara J\"ager
for valuable discussions. SD is supported by the United States
Department of Energy under Grant Contract DE-SC0012704 and is grateful
to the University of T\"ubingen, where this work was started. IML is
supported in part by United States Department of Energy grant number
DE-SC0017988. J.B. acknowledges the support from the Carl-Zeiss
foundation. Parts of this work were performed thanks to the support of
the State of Baden-W\"urttemberg through bwHPC and the DFG through the
grant no. INST 39/963-1 FUGG. The data to reproduce the plots has been
uploaded with the arXiv submission and is available upon request.
\end{acknowledgments}

\bibliographystyle{utphys}
\bibliography{ww}

\end{document}